\documentclass[twocolumn,aps,prl,floatfix,superscriptaddress]{revtex4-1}
\usepackage{amsmath,amssymb}
\usepackage{graphicx,float}
\usepackage{times,txfonts}
\usepackage{textcomp}
\usepackage{color,colortbl}
\usepackage{xcolor, dsfont, bbm}
\usepackage{multirow}
\usepackage{soul}
\usepackage{hyperref}
\usepackage{natbib}
\usepackage{nicefrac}
\usepackage{multirow}
\usepackage{blindtext}
\usepackage{makecell}
\usepackage[export]{adjustbox}
\definecolor{hugoColor}{RGB}{59,134,255}
\definecolor{Gray}{gray}{0.95}

\usepackage{tabularx}
\usepackage{upgreek}

\newcommand{\eqlab}[1]{\label{eq:#1}}
\renewcommand{\eqref}[1]{Eq.~(\ref{eq:#1})}

\newcommand{\figref}[1]{Fig.~\ref{fig:#1}}

\newcommand{\figlab}[1]{\label{fig:#1}}
\newcommand{\tabref}[1]{Table~\ref{tab:#1}}

\newcommand{\tablab}[1]{\label{tab:#1}}

\newcommand{\equal}{\!=\!}

\begin{document}

\title{Photonic crystal cavity IQ modulators in thin-film lithium niobate for coherent communications}

\author{Hugo~Larocque}
\email{hlarocqu@mit.edu}
\affiliation{Research Laboratory of Electronics, Massachusetts Institute of Technology, Cambridge, MA 02139, USA}

\author{Dashiell~L.P.~Vitullo}
\affiliation{DEVCOM Army Research Laboratory, Adelphi, MD 20783-1193, USA}

\author{Alexander~Sludds}
\affiliation{Research Laboratory of Electronics, Massachusetts Institute of Technology, Cambridge, MA 02139, USA}

\author{Hamed~Sattari}
\affiliation{Centre Suisse d’Electronique et de Microtechnique (CSEM), 2000 Neuch\^atel, Switzerland}

\author{Ian~Christen}
\affiliation{Research Laboratory of Electronics, Massachusetts Institute of Technology, Cambridge, MA 02139, USA}

\author{Gregory~Choong}
\affiliation{Centre Suisse d’Electronique et de Microtechnique (CSEM), 2000 Neuch\^atel, Switzerland}


\author{Ivan~Prieto}
\affiliation{Centre Suisse d’Electronique et de Microtechnique (CSEM), 2000 Neuch\^atel, Switzerland}

\author{Jacopo~Leo}
\affiliation{Centre Suisse d’Electronique et de Microtechnique (CSEM), 2000 Neuch\^atel, Switzerland}

\author{Homa~Zarebidaki}
\affiliation{Centre Suisse d’Electronique et de Microtechnique (CSEM), 2000 Neuch\^atel, Switzerland}

\author{Sanjaya Lohani}
\affiliation{Department of Electrical and Computer Engineering, University of Illinois - Chicago, Chicago, IL 60607, USA}

\author{Brian T. Kirby}
\affiliation{DEVCOM Army Research Laboratory, Adelphi, MD 20783-1193, USA}
\affiliation{Tulane University, New Orleans, LA 70118, USA}

\author{{\"O}ney O. Soykal}
\affiliation{DEVCOM Army Research Laboratory, Adelphi, MD 20783-1193, USA}

\author{Moe~Soltani}
\affiliation{Raytheon BBN Technologies, Cambridge, MA 02138, USA}

\author{Amir~H.~Ghadimi}
\affiliation{Centre Suisse d’Electronique et de Microtechnique (CSEM), 2000 Neuch\^atel, Switzerland}

\author{Dirk~Englund}
\affiliation{Research Laboratory of Electronics, Massachusetts Institute of Technology, Cambridge, MA 02139, USA}

\author{Mikkel~Heuck}
\email{mheu@dtu.dk}
\affiliation{Department of Electrical and Photonics Engineering, Technical University of Denmark, 2800 Lyngby, Denmark}

\begin{abstract}

Thin-Film Lithium Niobate (TFLN) is an emerging integrated photonic platform showing great promise due to its large second-order nonlinearity at microwave and optical frequencies~\cite{Boyd2008}, cryogenic compatibility~\cite{Lomonte:21}, large piezoelectric response~\cite{Maleki:09}, and low optical loss at visible~\cite{Desiatov:19} and near-infrared~\cite{Zhang:17} wavelengths. These properties enabled Mach-Zehnder interferometer-based devices to demonstrate amplitude-~\cite{Wang:18} and in-phase/quadrature (IQ)~\cite{Xu:22} modulation at voltage levels compatible with complementary metal-oxide-semiconductor (CMOS) electronics. Maintaining low-voltage operation requires centimeter-scale device lengths, making it challenging to realize the large-scale circuits required by ever-increasing bandwidth demands in data communications~\cite{Cheng:18}. Reduced device sizes reaching the 10~$\upmu$m scale are possible with photonic crystal (PhC) cavities. So far, their operation has been limited to modulation of amplitudes and required circulators~\cite{Li:20} or lacked cascadability~\cite{Prencipe:21}. Here, we demonstrate a compact IQ modulator using two PhC cavities operating as phase shifters in a Fabry-Perot-enhanced Michelson interferometer configuration~\cite{Shoemaker:91}. It supports cascadable~\cite{Xu:06} amplitude and phase modulation at GHz bandwidths with CMOS-compatible voltages. While the bandwidth limitation of resonant devices is often considered detrimental, their compactness enables dense co-integration with CMOS electronics where clock-rate-level operation (few GHz) removes power-hungry electrical time-multiplexing~\cite{Miller:17, Lee:23}. Recent demonstrations of chip-scale transceivers with dense-wavelength division multiplied transceivers~\cite{Rizzo:23} could be monolithically implemented and driven toward ultimate information densities using TFLN electro-optic frequency combs~\cite{Hu:22} and our PhC IQ modulators.



\end{abstract}

\maketitle

\section{Introduction}

Modern telecommunications rely on electro-optic (EO) modulators to convert information between electrical and optical signals~\cite{Sinatkas:21, Liu:04,Xu:05,Reed:10,Timurdogan:14,Han:17,Hiraki:17,Pintus:22,Xiong:12,Wang:18, Abel:19}. 
The exponentially increasing demand for information capacity~\cite{Winzer:17} and growing interest in networking superconducting quantum circuits~\cite{Xiang:13} motivates the development of small-footprint EO modulators with low power consumption that can be densely integrated with electronic processors~\cite{Miller:17} while operating near the fundamental limits given by the interaction between microwave and optical photons~\cite{Han:21}. Coherent communications have proven instrumental in leveraging existing technology for high bandwidth internet protocol optical routing~\cite{Rohde:14,Agrell:16} and enhancing throughput in long-haul fiber networks~\cite{Schmogrow:12, Pfeifle:14, Kikuchi:16, Wang:18_2}, while promising similar features for data center interconnects and edge computing~\cite{Cheng:18, Sludds:22}. 
\begin{figure*}[t]
	\centering
	\includegraphics[width=\linewidth]{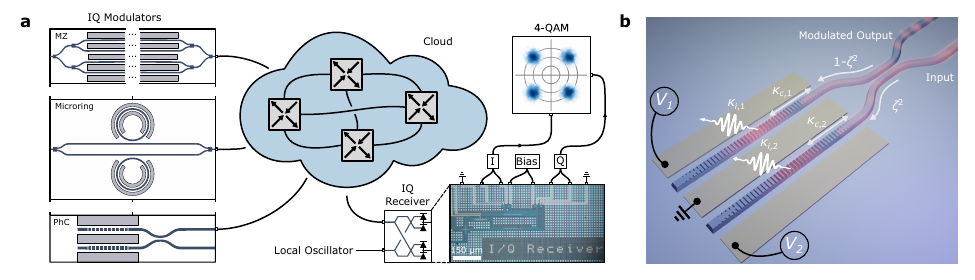}
	\caption{\textbf{Coherent communications with IQ modulators.} {\bf a,} Sketch of coherent communication network relying on both IQ modulators and receivers. PIC-based IQ modulators include hardware based on Mach-Zehnder (MZ) interferometers and more compact resonant structures, such as microring resonators and the photonic crystal-based device presented in this work. The dots in the MZ sketch highlight its extent over a scale exceeding that of its illustrated features. PIC-based IQ receivers are available through silicon photonics manufacturing, and the micrograph depicts the device used in this work. {\bf b,} Schematics of our coherent modulator based on one-sided photonic crystal cavities in a Michelson interferometer configuration. }
	\figlab{fig1}
\end{figure*}
As illustrated in~\figref{fig1}a, these advances in coherent communications hinge on in-phase/quadrature (IQ) modulators, which are able to control both the amplitude and phase of optical fields and are currently sustained by commercially available InP-based devices~\cite{Rouvalis:15,Ozaki:18}.

Further technological requirements have driven rapid advances in photonic integrated circuits (PICs) with EO modulators~\cite{Sinatkas:21} based on interactions including free-carrier dispersion~\cite{Liu:04,Xu:05,Reed:10,Timurdogan:14}, the quantum confined Stark effect~\cite{Han:17,Hiraki:17,Pintus:22}, and the Pockels effect~\cite{Xiong:12,Wang:18, Abel:19}. Advances in silicon photonics have notably enabled a new generation of coherent optical engines~\cite{Novack:18, Nokia:23} along with more compact implementations relying on microring phase shifters~\cite{Dong:12}. Such free carrier-based devices face fundamental trade-offs between insertion loss and modulation efficiency that ultimately cap their performance. The pure phase response of Pockels materials can overcome these challenges.
For example, thin-film lithium niobate (TFLN) is a promising PIC platform due to its wide transparency window, large Pockels coefficients $r_{ij}$, and low waveguide loss~\cite{Wang:18}. When arranged in a traveling-wave Mach-Zehnder configuration, TFLN modulators achieve modulation rates exceeding 100 GHz~\cite{Wang:18, Xu:20, Xu:22, Kharel:21} and can naturally integrate into IQ modulator architectures~\cite{Xu:20, Xu:22}. However, as emphasized in~\figref{fig1}a, their length needs to extend over several millimeters to reach sufficient microwave-to-optical interaction strengths, which could prevent their use in applications requiring high co-integration densities. The modulator size has been reduced using structures such as folded Michelson interferometers~\cite{Xu:19} and microring-assisted Mach-Zehnder interferometers~\cite{Feng:22,Menssen:23}. Dielectric photonic crystal (PhC) cavities provide wavelength-scale confinement without compromising insertion loss. 
As shown in a recent demonstration of off-keying in TFLN PhC cavities~\cite{Li:20}, this resonant modulation scheme preserves the alignment between LN's Pockels tensor and the modulating electric field over a device with an ultra-small capacitance and an optical mode volume as low as 0.58~\textmu m$^3$. However, it has remained an open challenge to develop devices with 2 degrees of freedom -- the minimum needed for arbitrary modulation of the two field quadratures. 

Here, we solve this challenge by introducing an ultrasmall TFLN PhC IQ modulator, taking advantage of wavelength-scale confinement through PhC cavities in an interferometric configuration. We demonstrate four symbol quadrature- and amplitude modulation (4-QAM) with a complementary-metal-oxide-semiconductor (CMOS) compatible peak-to-peak driving voltage of 2$\,$V. The modulation rate of $\sim \!\!1\,$GHz is limited by the cavity quality factor ($Q$) of $\sim \!\!70,000$, and our electrode configuration results in a tuning efficiency of $\sim \!\!1\,$GHz/V. Through iterative co-design and testing of cavity Bragg mirrors and stable fabrication process control in wafer-scale TFLN manufacturing, we achieve a fabrication yield exceeding 64\% for PhCs with $Q$ values above $20,000$ across devices with design parameters specified in the Supplementary.

\section{Results}
The conventional ring-resonator-enhanced Mach-Zehnder architecture~\cite{Dong:12, Feng:22, Menssen:23} does not carry over to PhC cavities, as they couple the incident field to forward- and backward-propagating waves. We therefore developed a different design: a cavity-assisted on-chip Michelson interferometer, inspired by laser interferometric gravitational wave detectors~\cite{Shoemaker:91} that use two arms of one-sided Fabry Perot cavities. In our design, a directional coupler with a $\zeta^2$:$1-\zeta^2$ splitting ratio distributes an input optical signal to two \textit{one-sided Fabry-Perot PhC cavities} where light couples to the waveguides at rates $\kappa_{c,1}$ and $\kappa_{c,2}$, see~\figref{fig1}b. 
%
\begin{figure*}[t]
	\centering
	\includegraphics[width=\linewidth]{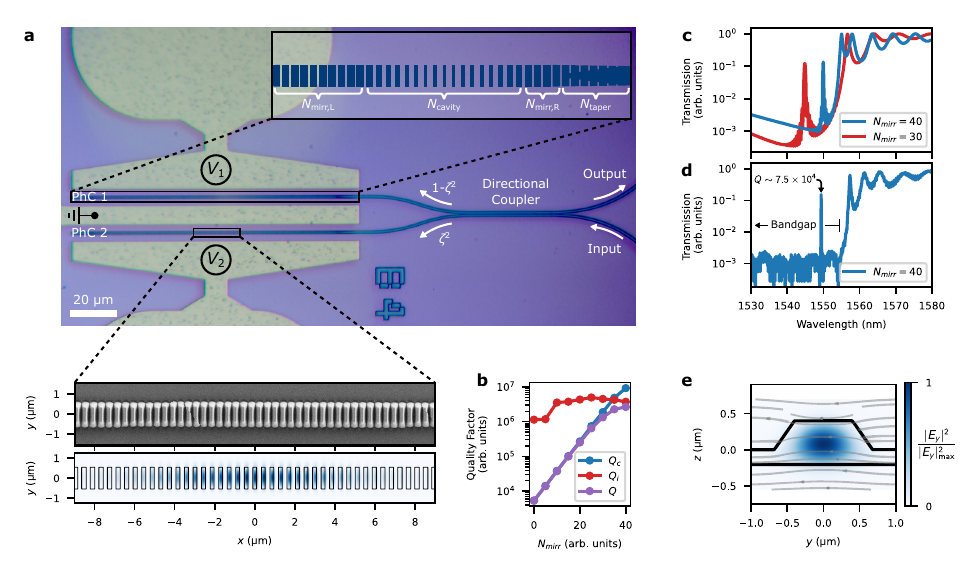}
    \caption{\textbf{Device fabrication and design.} {\bf a,} Optical micrograph of our PhC cavity IQ modulator. Top inset: Schematic of the cavity. Bottom inset: Scanning electron microscopy image of a PhC cavity and the corresponding optical mode calculated using FDTD simulations. {\bf b,} Coupling ($Q_c$), intrinsic ($Q_i$), and total ($Q$) quality factors of PhC cavities as a function of $N_{\rm{mirr}}\equal N_{\rm{mirr,L}}\equal N_{\rm{mirr,R}}$ obtained via FDTD simulations. {\bf c,} Simulated transmission spectra of a two-sided PhC cavity with $N_\text{mirr}\equal 40$ (blue), and a one-sided cavity with $N_\text{mirr,R}\equal 30$ and $N_\text{mirr,L}\equal 100$. {\bf d,} Measured transmission spectrum of a two-sided reference cavity with $N_\text{mirr}\equal 40$ corresponding to {\bf c}. {\bf e,} Finite element simulation of the DC electric field (streamlines) induced by an applied voltage with a density plot of the optical eigenmode overlaid.
    }\figlab{fig2}
\end{figure*}
Pairs of electrodes apply electric fields across the TFLN cavities by means of the voltages $V_1$ and $V_2$. To first order in $V_n$, the cavity resonance frequencies $\omega_n$ shift by (see Supplementary Section II)
\begin{eqnarray}\eqlab{tuning efficiency}
    \Delta\omega_n = \omega_n - \omega_n^{(0)} = \frac{\partial \omega_n}{\partial V}  V_n \equiv \partial_V\omega_n V_n, ~~~ n\equal\big(1,2\big),
\end{eqnarray}
where $\omega_n^{(0)}$ is the resonance at zero voltage, and $\partial_V\omega_n$ is the tuning efficiency. Cavity loss is described by the intrinsic decay rates $\kappa_{i,1}$ and $\kappa_{i,2}$. After reflection from the cavities, the modulated signals travel back across the directional coupler and interfere. The input-output transmission is (see Supplementary Section III)
\begin{subequations} \eqlab{transmission}
\begin{align}
    t_\text{IQ}(\omega) &= \zeta \sqrt{1-\zeta^2} \bigg(r_1(\omega)e^{i\Delta \phi} + r_2(\omega)\bigg), \eqlab{tIQ}\\
     r_n(\omega) &= 1 - \frac{2\kappa_{c,n}}{i2\delta_n + \kappa_{c,n} + \kappa_{i,n}}, \eqlab{rn}
\end{align}
\end{subequations}
where $r_n(\omega)$ is the cavity reflection coefficient, $\delta_n \equal \omega_n - \omega$ is the detuning from the input carrier frequency, $\kappa_n \equal \kappa_{c,n} + \kappa_{i,n}$ is the total linewidth, and $\Delta \phi$ describes the relative phase between the interferometer arms. The transmission $t_{\rm{IQ}}$ attains any complex value within the unit circle in the limit of highly over-coupled cavities ($\kappa_{c,n}\!\gg\!\kappa_{i,n}$) with a 50:50 directional coupler. The condition for complete extinction ($t_{\rm{IQ}}\equal 0$) is independent of the splitting ratio $\zeta$, which is not the case for Mach-Zehnder implementations containing two distinct couplers.

Figure~\ref{fig:fig2}a shows a micrograph of our fabricated TFLN IQ modulator. The top-right inset sketches the PhC cavity to illustrate its formation by modulating the width of a waveguide. As detailed in Supplementary Section IV, we vary the duty cycle parabolically over $N_{\rm{cavity}}$ periods to produce a high $Q$ resonance~\cite{Quan:09}. Placing fewer mirror periods on the side facing the waveguide ($N_{\rm{mirr,R}}\!<\!N_{\rm{mirr,L}}$) achieves a one-sided configuration. A smooth transition to the propagating waveguide mode minimizes out-of-plane scattering by linearly reducing the width modulation to zero over $N_{\rm{taper}}$ periods.
Figure~\ref{fig:fig2}b plots the total- (purple), coupling- (blue), and intrinsic (red) quality factors, calculated using finite-difference-time-domain (FDTD) simulations, as a function of the number of mirror periods $N_{\rm{mirr}}$ for a two-sided cavity ($N_{\rm{mirr,L}}\equal N_{\rm{mirr,R}}\equal N_{\rm{mirr}}$). It highlights how $Q$ is easily adjusted to match a targeted modulation speed without sacrificing the intrinsic quality factor $Q_{i,n}\equal\omega_n/\kappa_{i,n}$. The corresponding simulated transmission spectra are shown in~\figref{fig2}c, and the measured spectrum from a two-sided reference cavity is plotted in~\figref{fig2}d. The good agreement between measurement and simulation (blue curves in Figs.~\ref{fig:fig2}c,d) results from extracting geometrical information via scanning-electron-microscope images and additional reference structures (see Supplemental Section V for details).

We calculate a tuning efficiency of $\partial_V \omega_n \equal 2\pi \!\times\! 1.0$~GHz/V via first-order perturbation theory based on the overlap between the optical cavity mode and the field from the electrodes (see Supplementary Section VI).
Figure~\ref{fig:fig2}e shows how the electrode field (streamlines) and the optical field (blue contour) are parallel to maximize their overlap. 
\begin{figure}[t]
	\centering
	\includegraphics[width=\columnwidth]{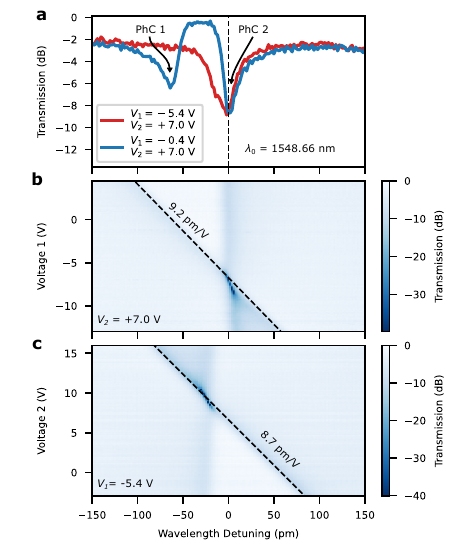}
	\caption{\textbf{Electro-optic resonance shifts.} {\bf a,} Transmission spectrum of the IQ modulator under bias voltages of $(V_1,V_2) = (-5.4~$V$,7.0~$V$)$ and $(V_1,V_2) = (-0.4~$V$,7.0~$V$)$. The first set of voltages (red) aligns the resonances of the cavities and is closer to the nominal bias used in the rest of this work, whereas the second configuration (blue) displays the individual cavity resonances. {\bf b,} Transmission spectrum of the modulator while sweeping the actuation voltages on the device's first and {\bf c,} second PhC cavities. The dashed black lines trace the resonance shifts determined by the resonant wavelengths, $\lambda^{(0)}$, and the tuning efficiencies, $\partial_V\omega$, from \tabref{fitted parameters table}. 
 }
 \figlab{fig3}
\end{figure}
Experimentally, we determine the tuning efficiency by measuring the transmission at different voltage settings. Figure~\ref{fig:fig3}a plots two spectra with the resonances aligned (red) or separated (blue). Maps of transmission versus frequency and voltage across one of the cavities are shown in Figs.~\ref{fig:fig3}b,c. The transmission dips caused by cavity resonances are observed to shift linearly in response to the applied voltage. As described in Supplementary Section VII, we fit the data from Figs.~\ref{fig:fig3}b,c to \eqref{transmission}, thereby obtaining the model parameters listed in~\tabref{fitted parameters table}.
%
\begin{figure*}[!h]
	\centering
 \includegraphics[width=\linewidth]{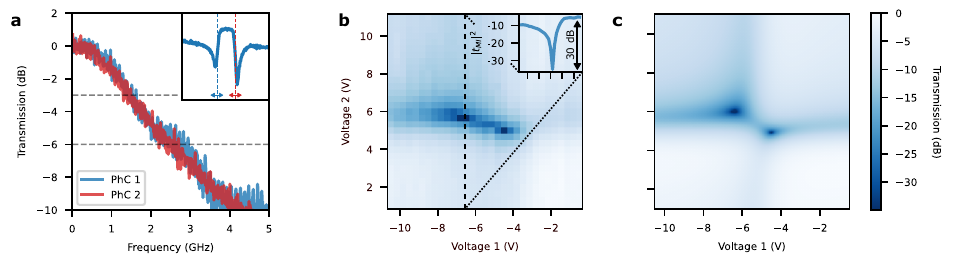}
	\caption{\textbf{Device bandwidth and extinction ratio measurements.} {\bf a,} Small signal response of each cavity as a function of the applied frequency. Inset: Dashed lines indicate the wavelengths used. {\bf b,} Transmission of the IQ modulator, $|t_\text{IQ}|^2$, as a function of the voltages $(V_1,V_2)$ applied to each cavity for an optical wavelength of $\lambda=1548.66$~nm. Inset: Transmission along the vertical dashed line. {\bf c,} Transmission calculated from~\eqref{tIQ} using fitted parameters for the same voltage interval as in {\bf b}. 
 }
    \figlab{fig4}
\end{figure*}

Figure~\ref{fig:fig4}a plots the small-signal modulator response when each PhC cavity is driven by a sinusoidal voltage signal. We choose the DC voltage offsets and laser wavelength to maximize the signal-to-noise of the transmitted light (see inset).
Each cavity has a 3~dB cutoff around 1.5~GHz, which matches well with the fitted decay rates listed in~\tabref{fitted parameters table}. 

%
\begin{table}[!h]
\renewcommand{\arraystretch}{1.4}
\vspace{3mm}
   \small
    \begin{tabular}{ |p{1.5cm}|p{2.5cm}|p{1.5cm}|p{2.5cm}|  }
    \hline
    \rowcolor{Gray}
     \multicolumn{2}{|c|}{\bf Cavity 1} & \multicolumn{2}{c|}{\bf Cavity 2} \\
     \hline
     \makecell{$\lambda_1^{(0)}$}  & \makecell{1548.60$\,$nm}  & \makecell{$\lambda_2^{(0)}$}    & \makecell{1548.71$\,$nm}  \\
     \hline
     \rowcolor{Gray}
     \makecell{$Q_1$} & \makecell{$6.6 \times 10^4$} & \makecell{$Q_2$} & \makecell{$7.4 \times 10^4$} \\
     \hline
     \makecell{$\kappa_1$} & \makecell{$2\pi \times  2.9$~GHz} & \makecell{$\kappa_2$} & \makecell{$2\pi \times  2.6$~GHz} \\
     \hline
     \rowcolor{Gray}
     \makecell{$\kappa_{c,1}/\kappa_{i,1}$} & \makecell{0.46} & \makecell{$\kappa_{c,2}/\kappa_{i,2}$} & \makecell{1.42} \\
     \hline
     \makecell{$\partial_V\omega_1$} & \makecell{$2\pi\times 1.15$~GHz/V} & \makecell{$\partial_V\omega_2$} & \makecell{$2\pi\times 1.09$~GHz/V} \\
     \hline
     \rowcolor{Gray}
     \multicolumn{4}{|c|}{\bf Michelson Interferometer}  \\
     \hline
     \makecell{$\Delta\phi$} & \makecell{$0.63\pi$} & \makecell{$\zeta$} & \makecell{$\sqrt{0.12}$} \\
     \hline
    \end{tabular}
    \caption{Model parameters extracted by fitting the measured data in~\figref{fig3}b,c and~\figref{fig4}b using~\eqref{tIQ}. }
    \tablab{fitted parameters table}
\end{table}

To better understand how to set DC bias voltages for QAM modulation, we measure the transmission as a function of both voltages at a fixed laser wavelength. The result is shown in~\figref{fig4}b, and~\figref{fig4}c plots the simulated transmission map from~\eqref{tIQ} using the parameters in~\tabref{fitted parameters table}. 
Destructive interference between the signals reflected from the two PhC cavities gives rise to the local minima exhibiting more than 30~dB extinction.  
%
\begin{figure*}[!h]
    \centering
    \includegraphics[width=\linewidth]{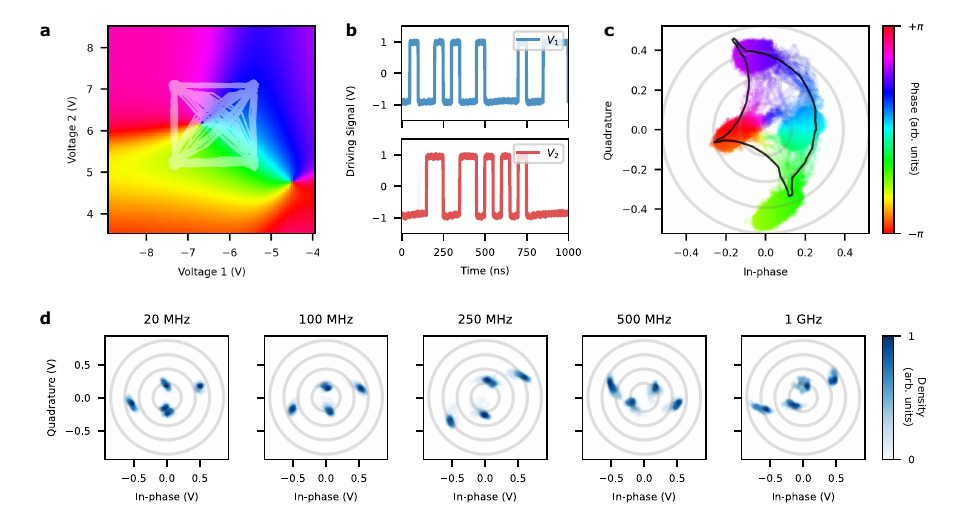}
    \caption{\textbf{IQ modulation.} {\bf a,} Modeled phase transmission map, $\arg\{t_\text{IQ}(V_1,V_2)\}$, obtained with the same model parameters as~\figref{fig4}c. The semi-transparent white lines plot the voltage settings of the RF driving signal. {\bf b,} Subset of 20~MHz pseudo-random bit sequences driving each PhC cavity, corresponding to the white traces in {\bf a}. {\bf c,} In-phase, $I$, and quadrature, $Q$, components of the modulated field measured using an IQ receiver. The black outline corresponds to the calculated outline formed by the trajectory in $(V_1, V_2)$ drawn in {\bf a} using~\eqref{tIQ}. {\bf d,} Constellation diagrams of the modulated field extracted by sampling the measured IQ time traces at the repetition rate of the driving bit sequences. The considered frequencies range from 20~MHz to 1~GHz. 
    }
    \figlab{fig5}
\end{figure*}

The good agreement between measurement and modeling allows us to use the transmission phase $\arg\{t_{\rm{IQ}}(V_1,V_2)\}$ calculated from~\eqref{tIQ} to set the DC-bias point at $(V_1,V_2) \equal (4.4\,$V$,-4.15\,$V) while applying a radio-frequency (RF) modulation of $\pm$1~V to each cavity. Figure~\ref{fig:fig5}a plots this phase map, and the RF voltages of a pseudo-random bit sequence with $2^{7}-1$ symbols are plotted with semi-transparent white lines. Notice the large phase variations in the region between the two singularity points corresponding to the transmission minima in~\figref{fig4}c. Separated minima are only possible when the cavities are sufficiently close to being over-coupled. 
Figure~\ref{fig:fig5}b plots the voltages of a subset of the applied bit sequence for an example with a 20~MHz repetition rate. We collect the IQ-modulated signal by a lensed fiber and detect it using a silicon-photonics integrated IQ receiver (see Supplementary Section VIII).
In~\figref{fig5}c, we plot the measured raw coherent transmission trace of a continuous wave (CW) input field modulated over a time span of 6.5~\textmu s. 
Sampling the clustered points in~\figref{fig5}c allows the reconstruction of the modulated field's constellation diagram. We opt for a 1~GHz sampling frequency instead of the repetition rate of the driving pulses to consider a consistent amount of samples across the full set of modulation frequencies. Figure~\ref{fig:fig5}d provides such diagrams for modulation frequencies of 20~MHz, 100~MHz, 250~MHz, 500~MHz, and 1~GHz. Our results feature good data clustering at four distinct symbols exhibiting error vector magnitudes below 0.27 (see Methods) up to driving frequencies approaching the 3~dB cutoff of our IQ modulator. As discussed in Supplementary Section IX, optimized symbol separation is possible with more advanced encoding to account for the nontrivial dependence of $t_{\rm{IQ}}$ on $V_1$ and $V_2$. Such optimization procedures can also determine minimum device metrics for running coherent modulation processes. For example, given the insertion loss of our device, we require PhC cavity quality factors of at least $Q\sim 2\times10^4$ to run the 4-QAM experiments shown in this work. This condition determined our reported cavity fabrication yield of 64\%. 

\section{Discussion}
The compact size of our IQ modulator allows its energy consumption to be limited by its capacitance. This is a key requirement for low energy information processing~\cite{Miller:17} based on attojoule optoelectronics that could benefit emerging applications in photonics-based edge computing and inference~\cite{Sludds:22}. As discussed in Supplementary Section IV, we estimate an average value of 25.8~fJ per bit, though it could be reduced below 1~fJ/bit by appropriate design modifications~\cite{Li:20}. Compact and energy-efficient modulators reopens the trade space comprised by multiplexing in the temporal, spatial, and spectral domains~\cite{Lee:23}. The moderate bandwidth of energy-efficient high-$Q$ resonant modulators need not be a drawback since operating at a few GHz avoids power-hungry tasks such as electronic serialization~\cite{Miller:17} as well as clock and data recovery. 
For instance, a recent demonstration used silicon microring resonators for amplitude-modulation of 32 wavelength channels generated from a single laser using a silicon-nitride Kerr comb~\cite{Rizzo:23}. Similar TFLN implementations could monolithically integrate electro-optic combs~\cite{Hu:22} and our PhC IQ modulators to reduce footprint further and eliminate chip-to-chip coupling loss. Importantly, our PhC IQ modulators are cascadable like rings~\cite{Rizzo:23} since the transmission approaches 1 away from the resonances when $\Delta\phi\equal 0$ and $\zeta^2\equal 1/2$. The TFLN platform also benefits from recently introduced components, such as on-chip lasers~\cite{Shams-Ansari:22, Guo:23, Snigirev:23}, amplifiers~\cite{Chen:21}, and isolators~\cite{Yu:22}. Compact multiport switches were also proposed based on one-sided PhC cavity phase modulators~\cite{Heuck:23}. 

Reducing the interaction volume of electro-optic coupling between optical and RF fields and TFLN's cryogenic compatibility~\cite{Lomonte:21} introduces new prospects for quantum computing and networking, especially between microwave and optical single photons. Current implementations rely on coupled racetrack cavities~\cite{McKenna:20, Holzgrafe:20} with footprints that could be reduced by several orders of magnitude by switching to PhC cavities. Electro-optic control over tightly confined optical cavity modes was proposed for nonlinear quantum information processing~\cite{Heuck:20a, Li:20a, Krastanov:21} and would similarly benefit systems with integrated quantum emitters~\cite{Yang:23}.   

Future work should focus on stabilizing the optical response of our devices.
Such considerations include minimizing transmission drifts due to photorefractive effects, which are known to be significant in TFLN cavities~\cite{Zhu:21}. Mitigation strategies include cladding removal~\cite{Xu:21b}, elevated operating temperature~\cite{Yariv:96, Rams:00}, and doping~\cite{Wang:18_3, Jankowski:20,Kong:20}. For classical interconnect applications with significant variations in operating temperature, feedback control loops will be necessary~\cite{Christen:22, Lee:23, Rizzo:23}. Machine learning-assisted state-estimation~\cite{Lohani:20,Danaci:21,Lohani:22, Lohani:23} could play a crucial role in stabilizing the modulator's transmission and replacing conventional discrete signal processing methods to address channel mixing in coherent communications. Future investigations should additionally include energy reductions by replacing ohmic heaters~\cite{Lee:23, Rizzo:23} with non-volatile tuning mechanisms, such as phase-change materials~\cite{Bente:23}, electro-mechanical effects~\cite{Jiang:20}, or laser annealing of oxides~\cite{Lee:08,Chen:11,Panuski:22}.


In summary, we introduced an ultra-compact PIC-based electro-optic IQ modulator. By incorporating a pair of tunable PhC cavities in TFLN integrated photonics, we demonstrated GHz-rate coherent modulation of an optical field using CMOS-compatible driving voltages and a footprint of 40-by-200~\textmu m$^2$. Further size reduction is straightforward~\cite{Li:20}, which will pave the way towards dense co-integrated CMOS electronics and optical IQ modulators for large-scale EO modulation. 

\section{Methods}

\noindent {\textbf{Device Fabrication}} We fabricated our chip in one of CSEM's TFLN multi-project fabrication runs based on a 600 nm thick $x$-cut TFLN on insulator wafer from NanoLN. We etch the LN waveguides and PhCs using an HSQ mask patterned with electron-beam lithography. The etch is configured to remove 400 nm of LN and result in waveguides with a $35^\text{o}$ sidewall angle with respect to the normal of the chip. Within the gaps of the PhC's Bragg mirrors, SEM imaging and modeling of measured transmission data reveal that the sidewall angle is closer to $17^\text{o}$ (see Supplemental Section V). We pattern 500~nm thick gold electrodes with a liftoff process. Waveguides are designed to have a width of 800~nm that tapers out to 900~nm once they reach the PhC region of the device. We use a 660~nm gap in our modulator's directional coupler.

\vspace{1 EM}

\noindent {\textbf{PhC Design Parameters}} {We set the Bragg period in our IQ modulator's PhC cavities to 426~nm and the number of Bragg periods in the input mirror to $N_\text{mirr,R}=30$. The duty cycle of the Bragg mirrors is 68\% and tapers up to 83\% at the cavity center. We provide further details related to this tapering in Supplementary Section IV. For the experimental transmission measurements of the two-sided cavity shown in Fig.~\ref{fig:fig2}f, we show the results of the fabricated cavity with parameters most similar to our IQ modulator device. Here, the number of mirror periods is $N_\text{mirr,L}=N_\text{mirr,R}=40$ and the duty cycle of the cavity region is 80\%, while the Bragg period and the duty cycle of the mirrors are the same.}

\vspace{1 EM}

\noindent {\textbf{Simulation parameters}} {As specified by the fabrication process, our simulations assume a 600~nm thick TFLN membrane with a 400~nm ridge and a sidewall angle of $35^\text{o}$ attributed to the sides of the waveguide. We set the sidewall angles in the gaps formed by the Bragg structure to $17^\text{o}$ as approximated from SEM imaging and modeling. We provide further details on how these geometric parameters affect the transmission of the cavities in the Supplementary. We performed all finite-difference-time-domain (FDTD) simulations provided in this work with Ansys's Lumerical tools. Bandgap wavelengths of infinite Bragg mirrors were simulated using MIT Photonic Bands (MPB). We performed all finite element method (FEM) simulations with COMSOL Multiphysics.}

\vspace{1 EM}

\noindent {\textbf{Error Vector Magnitude Calculation}} {We rely on the following definition of the error vector magnitude (EVM) for each symbol of a constellation diagram:
\begin{equation*}
    \text{EVM} =  \sqrt{\frac{1}{N}\displaystyle \sum_{n=0}^{N-1} \frac{(i_n-i_0)^2 + (q_n-q_0)^2}{i_0^2 + q_0^2}}
\end{equation*}
where $N$ is the number of acquired samples attributed to a symbol, $(i_n, q_n)$ corresponds to the measured quadratures of the samples, and $(i_0, q_0)$ are the expected quadrature values of the symbol. 
The reported values attributed to a single constellation diagram correspond to the average EVMs across all of the diagram's symbols.
}

\vspace{1 EM}


\bibliography{phcBib}

\vspace{5mm}
\noindent \textbf{Acknowledgements}
H. L. acknowledges the support of the Natural Sciences and Engineering Research Council of Canada (NSERC), the Army Research Laboratory (Awards W911NF2120099 and W911NF2220127), and the QISE-NET program of the NSF (NSF award DMR-1747426). M. H. acknowledges funding from Villum Fonden (QNET-NODES grant no. 37417).
The authors acknowledge Ryan Hamerly, Cole Brabec, Saumil Bandyopadhyay, Jane E. Heyes,  Mingxiao Li, and Usman A. Javid for useful discussions. 


\end{document}


\title{Supplementary Information for: \\
Photonic crystal cavity IQ modulators in thin-film lithium niobate for coherent communications}
\author{Larocque et al.}

\begin{abstract}
\end{abstract}

\maketitle

\section{List of Parameters}
%
\begin{table}[ht!]
\centering
\renewcommand{\arraystretch}{1.3}
\begin{tabular}{ |p{1cm}|p{10.5cm}|p{2cm}| }
\hline \rowcolor{Gray}
 \thead{\bf Symbol}  & \thead{\bf Description} & \thead{\bf Unit}\\
 \hline
 \makecell{$\zeta$} & \makecell{Amplitude transmission of directional coupler} & \makecell{1} \\
 \hline  \rowcolor{Gray}
 \makecell{$r_n$} & \makecell{Amplitude reflection from cavity $n$} & \makecell{1} \\
 \hline
 \makecell{$t_{\rm{IQ}}$} & \makecell{Amplitude input-output transmission of IQ modulator} & \makecell{1} \\
 \hline  \rowcolor{Gray}
 \makecell{$\Delta\phi$} & \makecell{Phase difference between each arm of the Michelson interferometer} & \makecell{rad} \\
 \hline
 \makecell{$\omega_n$} & \makecell{Resonance frequency of cavity $n$} & \makecell{rad$\times$s$^{-1}$} \\
 \hline  \rowcolor{Gray}
 \makecell{$\omega_n^{(0)}$} & \makecell{Resonance frequency of cavity $n$ at zero-voltage} & \makecell{rad$\times$s$^{-1}$} \\
 \hline
 \makecell{$\omega$} & \makecell{Carrier frequency of input light} & \makecell{rad$\times$s$^{-1}$} \\
 \hline  \rowcolor{Gray}
 \makecell{$\delta_n$} & \makecell{Detuning between input light and cavity $n$ resonance ($\delta_n=\omega-\omega_n$)} & \makecell{rad$\times$s$^{-1}$} \\
 \hline
 \makecell{$\kappa_{c,n}$} & \makecell{Cavity-waveguide coupling rate of cavity $n$} & \makecell{rad$\times$s$^{-1}$} \\
 \hline  \rowcolor{Gray}
 \makecell{$\kappa_{i,n}$} & \makecell{Intrinsic decay rate of cavity $n$} & \makecell{rad$\times$s$^{-1}$} \\
 \hline
 \makecell{$\kappa_{n}$} & \makecell{Total decay rate of cavity $n$ ($\kappa_n = \kappa_{c,n} + \kappa_{i,n}$)} & \makecell{rad$\times$s$^{-1}$} \\
 \hline  \rowcolor{Gray}
 \makecell{$Q_{c,n}$} & \makecell{Quality factor corresponding to waveguide coupling ($Q_{c,n} = \omega_n/\kappa_{c,n}$)} & \makecell{1} \\
 \hline 
 \makecell{$\partial_V\omega_n$} & \makecell{Cavity tuning efficiency ($\partial_V\omega_n = \partial \omega_n/\partial V$ )} & \makecell{rad$\times$s$^{-1}$V$^{-1}$} \\  \rowcolor{Gray}
 \hline
 \makecell{$V_n$} & \makecell{Voltage across cavity $n$} & \makecell{V} \\
 \hline
 \makecell{$a$} & \makecell{Period of PhC cavity unit cell} & \makecell{m} \\
 \hline  \rowcolor{Gray}
 \makecell{$w_w$} & \makecell{Width of TFLN waveguide (see~\figref{cavity parameters}b,c) } & \makecell{m} \\
 \hline
 \makecell{$d_m$} & \makecell{Width modulation of unit cell} & \makecell{1} \\
 \hline  \rowcolor{Gray}
 \makecell{$w_m$} & \makecell{Inside width of modulated of unit cell, $w_m\equal w_w(1-d_m)$. } & \makecell{m} \\
 \hline
 \makecell{$d_a$} & \makecell{Duty cycle of unit cell} & \makecell{1} \\
 \hline  \rowcolor{Gray}
 \makecell{$l_a$} & \makecell{TFLN length within unit cell, $l_a=a(1-d_a)$, (see~\figref{cavity parameters}b,d)} & \makecell{m} \\
 \hline
 \makecell{$N_{\rm{cav}}$} & \makecell{Number of unit cells in the cavity section of the PhC cavity (see Fig. 2a)} & \makecell{1} \\
 \hline  \rowcolor{Gray}
 \makecell{$N_{\rm{mirr}}$} & \makecell{Number of unit cells in the mirror section of the PhC cavity (see Fig. 2a)} & \makecell{1} \\
 \hline
 \makecell{$N_{\rm{tap}}$} & \makecell{Number of unit cells in the taper section of the PhC cavity (see Fig. 2a)} & \makecell{1} \\  \rowcolor{Gray}
 \hline
 \makecell{$\theta_{o}$} & \makecell{Angle of the outer sidewall of the PhC waveguide (see~\figref{cavity parameters}c)} & \makecell{degrees} \\
 \hline
 \makecell{$\theta_{i}$} & \makecell{Angle of the inner sidewall of the PhC waveguide (see~\figref{cavity parameters}d)} & \makecell{degrees} \\
 \hline  \rowcolor{Gray}
 \makecell{$h_w$} & \makecell{Thickness of TFLN waveguide (see~\figref{cavity parameters}d)} & \makecell{m} \\
 \hline
 \makecell{$h_m$} & \makecell{Thickness of TFLN membrane (see~\figref{cavity parameters}d)} & \makecell{m} \\
 \hline \rowcolor{Gray}
 \makecell{$l_0$} & \makecell{Coupling length of the directional coupler (see~\figref{directional_coupler}a)} & \makecell{m} \\
 \hline 
 \makecell{$d_0$} & \makecell{Waveguide separation of the directional coupler (see~\figref{directional_coupler}a)} & \makecell{m} \\
 \hline 
\end{tabular}
\caption{\textbf{Parameter descriptions.}}
\label{table:symbols table}
\end{table}

\section{Perturbation Theory \seclab{Perturbation Theory}}
Maxwell's equations for linear, anisotropic media without magnetization, free currents, and free charges are
\begin{subequations}\eqlab{Maxwells equations}
\begin{align}
    \nabla \times \vec{\mathcal{E}} &= -\frac{\partial \vec{\mathcal{B}}}{\partial t} \\
    \nabla \times \vec{H} &= \frac{\partial \vec{\mathcal{D}}}{\partial t} \\
    \nabla\cdot \vec{\mathcal{B}} &=0 \\
    \nabla\cdot \vec{\mathcal{D}} &=0 \\
    \vec{H}           &= \frac{1}{\mu_0} \vec{\mathcal{B}} \quad \text{and} \quad \vec{\mathcal{D}} = \epsilon_0\vec{\mathcal{E}} + \vec{\mathcal{P}}. \eqlab{constitutive relations}
\end{align}
\end{subequations}
For nonlinear media, we have the general expression for the polarization field
\begin{align}\eqlab{P-field definition} 
    \vec{\mathcal{P}} &= \epsilon_0 \left (  \chi^{(1)}\!:\! \vec{\mathcal{E}} + \chi^{(2)}\!:\!\vec{\mathcal{E}}\vec{\mathcal{E}} + \chi^{(3)}\!:\!\vec{\mathcal{E}}\vec{\mathcal{E}}\vec{\mathcal{E}} \right) + \mathcal{O}\big(|\vec{\mathcal{E}}|^4\big). 
\end{align}
The linear contribution is written in terms of the permittivity matrix
\begin{align}\eqlab{eps_r}
   \vec{\mathcal{P}}^{(1)} = \epsilon_0\bm{\epsilon}_r(\vec{r}) \vec{\mathcal{E}} \quad \text{with} \quad   \bm{\epsilon}_r(\vec{r}) = \bm{\mathbb{I}} + \bm{\chi}^{(1)}(\vec{r}),
\end{align}
where $\bm{\mathbb{I}}$ is the identity matrix. In our case, the second-order contribution to the polarization field describes the Pockels effect, where a DC field at $\omega_{\rm{DC}}\approx 0$ interacts with an optical cavity field at $\omega_n$. The resulting polarization field is~\cite{Boyd2008}
\begin{align}\eqlab{chi-2 P}
    \hat{\mathcal{P}}_i^{(2)}(\vec{r},\omega_n) = \epsilon_0 \sum_{j,k} \chi^{(2)}_{ijk}(\vec{r},\omega_n, \omega_{\rm{DC}},\omega_n) E_{{\rm{DC}},j}(\vec{r}) \hat{\mathcal{E}}_k(\vec{r},\omega_n) \equiv  \epsilon_0 \sum_k \Delta\epsilon_{i,k}^{(2)}(\vec{r}) \hat{\mathcal{E}}_k(\vec{r},\omega_n),
\end{align}
where $\vec{E}_{{\rm{DC}}}$ is the real-valued DC field and the permittivity perturbation matrix is defined as
\begin{align}\eqlab{Delta epsilon_r}
    \Delta\epsilon_{i,k}^{(2)}(\vec{r},\omega_n, \omega_{\rm{DC}},\omega_n) \equiv  \sum_j \chi^{(2)}_{ijk}(\vec{r},\omega_n, \omega_{\rm{DC}},\omega_n)E_{\rm{DC},j}(\vec{r}) .
\end{align}
The optical fields are written as expansions in eigenmodes
\begin{subequations}\eqlab{Field definitions 1}
\begin{align}
    \vec{\mathcal{E}}(\vec{r},t) &= \sum_{n} \frac12 \left( a_n(t) \vec{E}_n(\vec{r})e^{-i\omega_n t} + a_n^*(t) \vec{E}^{*}_n(\vec{r})e^{i\omega_n t}\right) \eqlab{E-field definition} \\
    \vec{\mathcal{B}}(\vec{r},t) &= \sum_{n} \frac12 \left( a_n(t) \vec{B}_n(\vec{r})e^{-i\omega_n t} + a_n^*(t) \vec{B}^{*}_n(\vec{r})e^{i\omega_n t}\right), \eqlab{B-field definition}
\end{align}
\end{subequations}
where the complex-valued mode functions, $\vec{E}_n$, obey the wave equation
\begin{align}\eqlab{wave equation} 
    \nabla\times \nabla \times \vec{E}_n - \mu_0\epsilon_0 \omega_n^2 \bm{\epsilon}_r \vec{E}_n &= 0.
\end{align}
From perturbation theory, the first-order correction to the eigenvalue is~\cite{Joannopoulos2008} 
\begin{align}\eqlab{first-order resonance shift a}
    \Delta\omega_n = \omega_n - \omega_n^{(0)} = -\frac12 \omega_n  \frac{\displaystyle \!\!\int \!\!dV  \Big(\bm{\Delta\epsilon}^{(2)} \,\vec{E}_n\Big)\cdot \vec{E}_n^*}{\displaystyle \int \!\!dV  \big(\bm{\epsilon}_r \vec{E}_n\big)\!\cdot\!\vec{E}_n^*} , 
\end{align}
where $\omega_n^{(0)}$ is the eigenvalue in the absence of a perturbation. We note here that the correct boundary conditions to use for~\eqref{wave equation} are radiation boundary conditions since the cavity modes will have finite quality factors~\cite{Kristensen2020}. Such leaky modes have complex eigenfrequencies, but in~\eqref{first-order resonance shift a} they are assumed to be real-valued. Mode functions, $\vec{E}_n(\vec{r},\omega_n)$, calculated using numerical methods that implement open boundary conditions e.g. via perfectly matched layers, will be leaky modes. This means that care must be taken when evaluating the normalization integral in the denominator of~\eqref{first-order resonance shift a} since these modes diverge for $|\vec{r}|\rightarrow\infty$. Typically, for high-quality factor, $Q_n$, cavities, there is a region of space surrounding the cavity where the field amplitude, $|\vec{E}_n|$, will be small and setting the boundaries of the normalization integral in this region provides reasonable results~\cite{Kristensen2020}.\\

We define the tuning efficiency $\partial_V\omega_n$ via the relation 
\begin{align}\eqlab{tuning efficiency}
    \Delta\omega_n = \omega_n - \omega_n^{(0)} = \frac{\partial \omega_n}{\partial V}  V_{\rm{DC}} \equiv \partial_V\omega_n V_{\rm{DC}},
\end{align}
where $\omega_n^{(0)}$ is the unperturbed eigenvalue and $V_{\rm{DC}}$ is the DC-voltage across cavity $n$. To arrive at an expression for $\partial_V\omega_n$ and define an interaction volume for the Pockels effect, we use the electrostatic relation between capacitance, DC voltage, and DC electric field
\begin{align}\eqlab{E DC normalization}
    \frac12 C V_{\rm{DC}}^2 = \frac12 \epsilon_0 \!\!\int \!\!dV  \big(\bm{\epsilon}_r \vec{E}_{\rm{DC}}\big)\!\cdot\!\vec{E}_{\rm{DC}}.
\end{align}
Inserting~\eqref{E DC normalization} into~\eqref{first-order resonance shift a}, we have
\begin{align}\eqlab{first-order resonance shift b}
    \Delta\omega_n & = -\frac12 \omega_n \sqrt{C} V_{\rm{DC}} \frac{\displaystyle  \!\!\int \!\!dV  \Big(\bm{\Delta\epsilon}^{(2)} \,\vec{E}_n\Big)\cdot \vec{E}_n^*}{\displaystyle \sqrt{  \epsilon_0 \!\!\int \!\!dV  \big(\bm{\epsilon}_r \vec{E}_{\rm{DC}}\big)\!\cdot\!\vec{E}_{\rm{DC}}} \int \!\!dV  \big(\bm{\epsilon}_r \vec{E}_n\big)\!\cdot\!\vec{E}_n^*}. 
\end{align}
Since $\bm{\Delta\epsilon}^{(2)}$ is proportional to the amplitude of the DC field, we see that the ratio of field integrals is independent of any field amplitudes, and this allows us to define the interaction volume as
\begin{align}\eqlab{V_EO definition}
    V_{\rm{EO}}     = \left( \frac{\displaystyle \sqrt{ \int \!\!dV  \Big(\bm{\epsilon}_r(\vec{r},\omega_{\rm{DC}}) \vec{E}_{\rm{DC}}(\vec{r})\Big)\!\cdot\!\vec{E}_{\rm{DC}}(\vec{r})} \!\int \!\!dV  \Big(\bm{\epsilon}_r(\vec{r},\omega_n) \vec{E}_n(\vec{r})\Big)\!\cdot\!\vec{E}_n^*(\vec{r})}{\displaystyle \int \!\!dV  \Big(\overline{\bm{\Delta\epsilon}}^{(2)}(\vec{r},\omega_n,\omega_{\rm{DC}},\omega_n) \,\vec{E}_n(\vec{r})\Big)\cdot \vec{E}_n^*(\vec{r})} \right)^2.
\end{align}
In~\eqref{V_EO definition}, we defined the normalized matrix 
\begin{align}\eqlab{normalized Delta epsilon}
    \overline{\bm{\Delta\epsilon}}^{(2)} = \frac{1}{r_{i,k}^{\rm{max}}} \bm{\Delta\epsilon}^{(2)},
\end{align}
where $r_{i,k}^{\rm{max}}$ is the maximum value of the Pockels coefficients in $\bm{\Delta\epsilon}^{(2)}$. Inserting~\eqref{V_EO definition} into~\eqref{first-order resonance shift b} and comparing to~\eqref{tuning efficiency}, we find
%
\begin{align}\eqlab{tuning efficiency b} 
    \partial_V\omega_n &= \frac12 \omega_n r_{i,k}^{\rm{max}} \sqrt{\frac{C}{\epsilon_0 V_{\rm{EO}}}}.
\end{align}

In our system, the DC field is directed along the extraordinary axis of the lithium niobate crystal, so we have~\cite{yariv2007}
\begin{align}\eqlab{Delta epsilon_r values}
    \bm{\Delta\epsilon}^{(2)} = -\left[\! \begin{array}{ccc} r_{13} \epsilon_o^2 & 0 & 0 \\
    0 & r_{13} \epsilon_o^2 &0 \\
    0 &0 & r_{33} \epsilon_e^2  \end{array} \!\right] E_{\rm{DC},3}, \quad \text{and} \quad   \bm{\epsilon}_r = \left[\! \begin{array}{ccc} \epsilon_o & 0 & 0 \\
    0 & \epsilon_o &0 \\
    0 &0 & \epsilon_e  \end{array} \!\right].
\end{align}
In all the coordinate systems used in this manuscript, the extraordinary axis is along the $y$-direction. Therefore, inserting~\eqref{Delta epsilon_r values} into~\eqref{first-order resonance shift a}, we get
\begin{align}\eqlab{resonance shift simple b} 
    \Delta\omega_n = \frac12 \omega_n  \frac{\displaystyle  \!\!\int \!\!dV  E_{\rm{DC},y} \bigg[r_{13}\epsilon_o^2\Big( |E_{n,x}|^2 \!+\! |E_{n,z}|^2\Big) + r_{33}\epsilon_e^2 |E_{n,y}|^2\bigg]}{\displaystyle \int \!\!dV  \bigg[ \epsilon_o \Big( |E_{n,x}|^2 \!+\! |E_{n,z}|^2\Big) + \epsilon_e |E_{n,y}|^2\bigg]}. 
\end{align}
%

\section{Temporal Coupled Mode Theory Modeling \seclab{Coupled Mode Theory Modeling}}
The input-output relation of the modulator is
\begin{align}\eqlab{io def T}
  \vec{E}_o = \left[\!\begin{array}{c} E_{o,1}\\ E_{o,2} \end{array} \!\right] = \bm{T} \vec{E}_i = \bm{T} \left[\!\begin{array}{c} E_{i,1}\\ E_{i,2} \end{array} \!\right],
\end{align} 
where the first subscript denotes inputs ($i$) and outputs ($o$), and the second subscript enumerates the upper and lower waveguide in Fig. 2a of the main text. The matrix, $\bm{T}$, is
\begin{align}\eqlab{def matrix M, R}
  \bm{T} = \bm{M}_{\rm{dc}}^{\rm{T}} \bm{M}_{\rm{mi}} \bm{R} \bm{M}_{\rm{mi}} \bm{M}_{\rm{dc}},
\end{align} 
with 
\begin{align}\eqlab{matrix T}
  \bm{M}_{\rm{dc}} = \left[\!\begin{array}{cc} \zeta & i\sqrt{1-\zeta^2} \\ i\sqrt{1-\zeta^2} & \zeta \end{array} \right], ~~ \bm{M}_{\rm{mi}} = \left[\!\begin{array}{cc} e^{i\Delta\phi/2} & 0 \\ 0 & 1 \end{array} \right], ~\text{and}~~ \bm{R} = \left[\!\begin{array}{cc} r_1(\omega) & 0 \\ 0 & r_2(\omega) \end{array} \right].
\end{align} 
The directional coupler is described by $\bm{M}_{\rm{dc}}$, the phase difference between the arms connecting the PhC cavities and the directional coupler is $\Delta\phi$, and the reflection coefficients of the PhC cavities are $r_n$ with $n\equal (1,2)$. The relevant transmission from the lower- to the upper waveguide is
\begin{align}\eqlab{tIQ}
  t_{\rm{IQ}}(\omega) \equiv -iT_{12}(\omega) = \zeta \!\sqrt{1-\zeta^2} \bigg(r_1(\omega)e^{i\Delta \phi} + r_2(\omega)\bigg).
\end{align} 
The reflection coefficients are found from the coupled mode theory description of each PhC cavity~\cite{Joannopoulos2008}
\begin{subequations}\eqlab{cmt cavities}
\begin{align}
  \frac{d}{dt} a_n &= \bigg(\!-i\omega_n - \frac{\kappa_{c,n}}{2} - \frac{\kappa_{i,n}}{2} \bigg)\,a_n - \sqrt{\kappa_{c,n}} b_{i,n} \\
  b_{o,n}   &= b_{i,n} + \sqrt{\kappa_{c,n}} a_n.
\end{align} 
\end{subequations}
If the input field oscillates at a carrier frequency, $\omega$, we define slowly varying envelopes as $a_n(t)=A_n(t) e^{-i\omega t}$ and $b_{i/o,n}(t)=B_{i/o,n}(t) e^{-i\omega t}$. The reflection spectrum is found by substituting the slowly varying amplitudes into~\eqref{cmt cavities} and setting $dA_n/dt=0$
\begin{subequations}\eqlab{cmt cavities spectrum}
\begin{align}
  0 &= \bigg(-i(\omega_n-\omega) - \frac{\kappa_{c,n}}{2} - \frac{\kappa_{i,n}}{2} \bigg) \,A_n - \sqrt{\kappa_{c,n}} B_{i,n} ~~\Rightarrow~~ A_n = -\frac{\sqrt{\kappa_{c,n}}}{i\delta_n + \kappa_n/2} \\
  r_n(\omega) &= |r_n(\omega)|e^{i\theta_n(\omega)} = \frac{B_{o,n}}{B_{i,n}}  = 1 - \frac{2\kappa_{c,n}}{ i2\delta_n + \kappa_n}, \eqlab{r_n spectrum}
\end{align} 
\end{subequations}
where the detuning is $\delta_n\equal \omega_n-\omega$ and the total cavity decay rate is $\kappa_n \equal \kappa_{c,n} + \kappa_{i,n}$.

For a 50/50 directional coupler ($\zeta\equal 1/\sqrt{2}$) and an infinite intrinsic quality factor ( $Q_{i,n}\equal \omega_n/\kappa_{i,n} \!\rightarrow\! \infty$), the cavity reflection is a pure phase response ($|r_n(\omega)| \equal 1$) and the transmission, $t_{\rm{IQ}}(\omega)$, attains all values within the unit circle when the resonance frequencies, $\omega_n$, are varied. Importantly, a zero transmission is attainable for arbitrary splitting ratio, $\zeta$, and quality factors. In the symmetric case, $\Delta\phi\equal 0$, there is an analytical solution for the two sets of resonance frequencies resulting in a zero transmission
\begin{align} \eqlab{zero transmission detunings}
    \delta_{1,\mp} = \mp \frac{1}{2} \sqrt{\kappa_{i,1} \left( \frac{\kappa_{c,1}\kappa_{c,2}}{\kappa_{i,2}} - \kappa_{i,1}\right)}, \qquad 
     \delta_{2,\pm} = \pm \frac{1}{2} \sqrt{\kappa_{i,2} \left( \frac{\kappa_{c,1}\kappa_{c,2}}{\kappa_{i,1}} - \kappa_{i,2}\right)}, \qquad \kappa_{1,c}\kappa_{2,c} \geq \kappa_{1,i}\kappa_{2,i}.
\end{align} 
The phase of the reflection, $\theta_n(\omega)$, from an over-coupled ($\kappa_{c,n}\!>\!\kappa_{i,n}$) one-sided cavity changes by $2\pi$ when the frequency is scanned across the resonance, $\omega_n$. Similarly, we see from~\eqref{zero transmission detunings} that there is a generalized requirement to observe two zero-transmission points. In~\figref{cmt examples}c, we plot the phase of $t_{\rm{IQ}}$ for two identical cavities with $\kappa_{c,n}\equal 6\kappa_{i,n}$.
\begin{figure*}[h!]
	\centering
\includegraphics[width=0.99\linewidth]{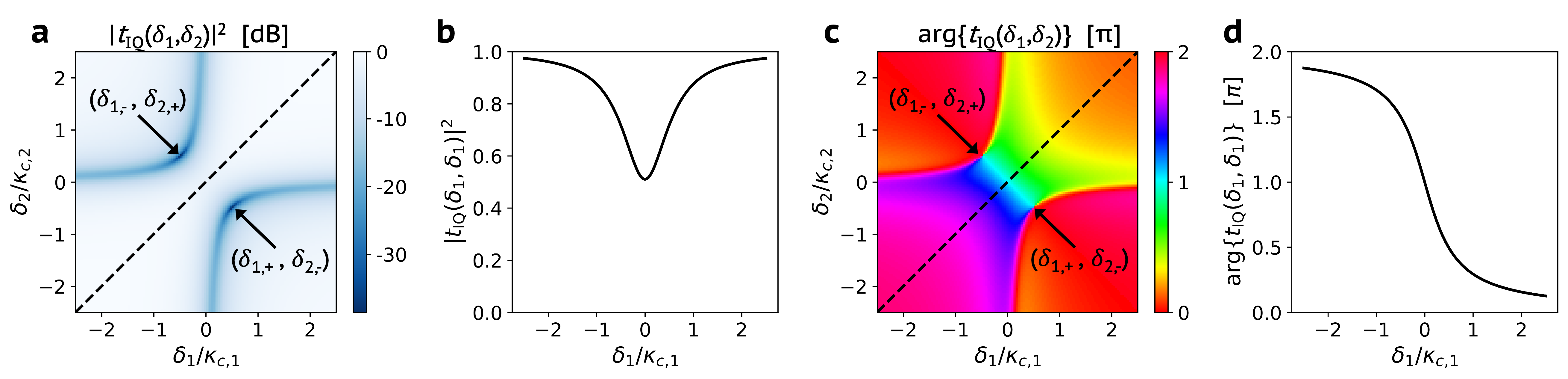}
	\caption{\textbf{Example of IQ modulator transmission.} {\bf a,} Intensity response, $10\log_{10}\{|t_{\rm{IQ}}(\delta_1, \delta_2)|^2\}$, using two identical PhC cavities. The zero-transmission solutions from~\eqref{zero transmission detunings} are indicated by black arrows. {\bf b,} Intensity response along the dashed black line in {\bf a}, which corresponds to the reflection from each cavity, $|r_n(\omega)|^2$. {\bf c,} Phase response, $\arg\{t_{\rm{IQ}}(\delta_1, \delta_2)\}$. The zero-transmission solutions from~\eqref{zero transmission detunings} are indicated by black arrows. {\bf d,} Phase response along the dashed black line in {\bf c}. In all plots, we used identical parameters for the two cavities with $\kappa_{c,n}\equal 6\kappa_{i,n}$, $\Delta\phi\equal 0$, and $\zeta\equal1/\sqrt{2}$. }
    \figlab{cmt examples}
\end{figure*}
Note that large phase variations are observed in the region between the zero-transmission points near the dashed black line along $\delta_2\equal \delta_1$. On this line, the transmission equals that of each cavity, $t_{\rm{IQ}}(\delta_1,\delta_1)\equal r_n(\delta_n)$, and the corresponding amplitude and phase are plotted in~\figref{cmt examples}b,d. For completeness, we also show the amplitude of the transmission in~\figref{cmt examples}a.

\section{Device Parameters}
\subsection{Photonic Chrystal Cavity}
The two PhC cavities used for the IQ-modulation experiments are nominally identical and formed by modulating the width, $w_w$, of a partially etched waveguide.~\figref{cavity parameters}a sketches the right part of the cavity that is divided into a ``cavity", ``right mirror", and ``taper" section each consisting of $N_{\rm{cavity}}$, $N_{\rm{mirr,R}}$, and $N_{\rm{taper}}$ periods.~\figref{cavity parameters}b-d show schematic cross sections of a single unit cell in the taper section, where a straight waveguide gradually transforms into a photonic crystal Bragg mirror.
\begin{figure*}[h!]
	\centering
\includegraphics[width=0.8\linewidth]{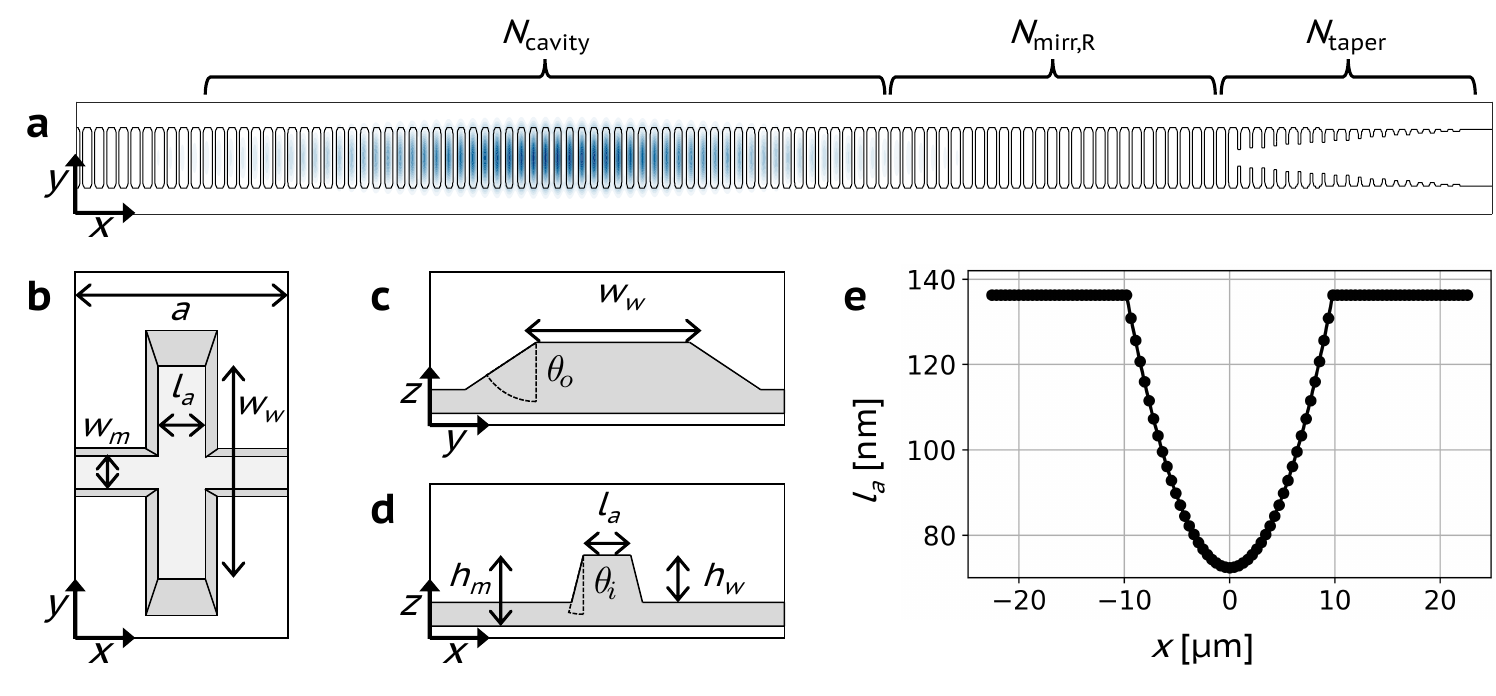}
	\caption{\textbf{PhC cavity parameters.} {\bf a,} Top-view $x,y$ cross section indicating the different sidewall angles in the $x$- and $y$-directions. The cavity-, right mirror-, and taper sections are indicated. {\bf b,} Sketch of the rightmost part of a PhC cavity with the optical mode overlaid in blue. {\bf c,} Side-view $y,z$ cross section. {\bf d,} Side-view $x,z$ cross section. {\bf e,} Variation of $l_a$, as sketched out in {\bf b}, with $x$, where $x\equal 0$ corresponds to the center of the cavity.}
    \figlab{cavity parameters}
\end{figure*}
The width modulation, $d_m\equal (w_w-w_m)/w_w$, is varied linearly over $N_{\rm{taper}}\equal 24$ periods from 0 at the straight waveguide to 1 at the right mirror section. The waveguide width, $w_w$, also changes linearly from 800$\,$nm to 900$\,$nm. In the mirror and cavity sections, each segment has a length of $l_a\equal a(1-d_a)$, where $d_a$ is the duty cycle and $a=426\,$nm is the PhC period.~\figref{cavity parameters}e shows how the duty cycle changes parabolically from $d_a\equal0.83$ to $d_a\equal0.68$ over $N_{\rm{cavity}}\equal 24$ periods and stays constant in the right mirror section for $N_{\rm{mirr,R}}\equal 30$ periods. Note that~\figref{cavity parameters}e shows a symmetric two-sided cavity configuration, whereas the cavities in the IQ modulator are asymmetric one-sided cavities with $N_{\rm{mirr,L}}\equal 100$. 
The sidewall angle is significantly smaller ($\theta_i<\theta_o$) in the $x$-direction than the $y$-direction due to smaller distances between etched features, see~\figref{cavity parameters}c,d and explanations below.  
\subsection{Directional Coupler}
The directional coupler ensuring the interferometric coupling between the modulated signals of each cavity has a nominal length of $l_0\equal 34.5$~\textmu m and waveguide separation of $d_0\equal$ 660~nm, see~\figref{directional_coupler}a. The waveguide width, $w_w$, is tapered down from the nominal 800~nm to $w_0\equal 600~$nm in the coupling region.~\figref{directional_coupler}b plots the measured transmission of the coupler in its bar and cross ports, indicating that its power splitting ratio at the operation wavelength of 1548.65~nm is near 12:88. 
\begin{figure*}[h]
	\centering
\includegraphics[width=0.9\linewidth]
{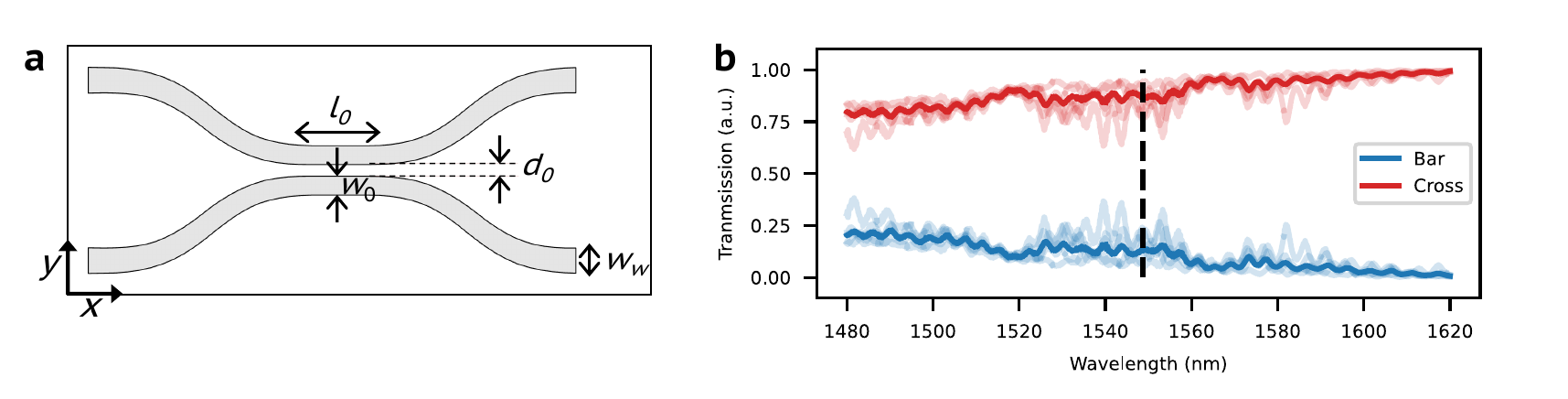}
	\caption{\textbf{Directional coupler splitting ratio.} {\bf a,} Sketch of the coupler indicating geometrical parameters. {\bf b,} Measured wavelength dependence of the transmission in the bar- and cross ports of the directional coupler used in the TFLN PhC IQ modulator.}
    \figlab{directional_coupler}
\end{figure*}
Measured transmission spectra from 5 different devices with the same nominal design are plotted with semi-transparent lines, and the solid lines show the averages.

\section{Iterative fabrication for PhCs with bandgaps in the C-band}
We fabricated TFLN Bragg waveguide mirrors to inform the design of the PhC cavities in the IQ modulator. The first round of fabrication considered structures of varying width, period, and duty cycle. Scanning electron microscope (SEM) images of the devices indicate clear differences between the $\theta_o\equal 35^\text{o}$ outer sidewall angle of the edge of the waveguides (see Methods) and the inner sidewall angle, $\theta_i$, in the gaps of the Bragg mirrors, see~\figref{cavity parameters}b-d.~\figref{semAngle} provides SEM images of a Bragg mirror while the sample is tilted at different angles in the plane formed by the direction parallel to the waveguide and the normal of the chip's surface. In these images, tilting the sample between an angle of $15^\text{o}$ and $20^\text{o}$ hides one of the two inner sidewalls formed by the gaps, thereby indicating that the inner sidewall angle lies within this range.
\begin{figure*}[h]
	\centering
	\includegraphics[width=\linewidth]{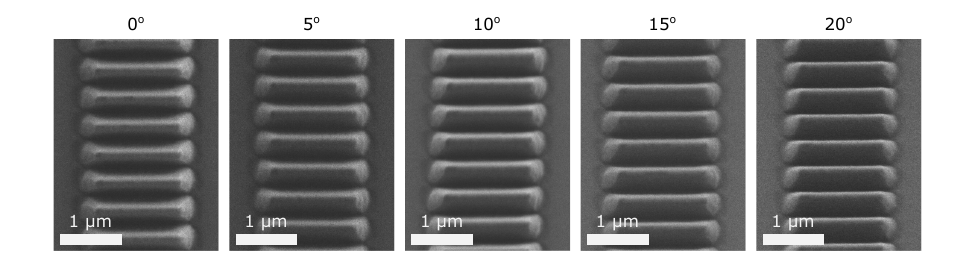}
	\caption{\textbf{Tilted SEM imaging of TFLN Bragg gratings.} SEM images of a TFLN Bragg mirror with a width of $w_w\equal1500$~nm, a duty cycle of $d_a\equal0.8$, and a period of $a\equal 450$~nm while the device is tilted by angles ranging from $0^\text{o}$ to $20^\text{o}$.}
        \figlab{semAngle}
\end{figure*}

~\figref{round1Trans} plots the transmission spectrum of Fabry Perot cavities formed by two Bragg mirrors and a 42~\textmu m cavity region consisting of a straight waveguide. We relied on the apparatus presented in~\secref{transApparatus} to collect these spectra, which clearly shows a red shift of the edge of a bandgap as the period of the Bragg mirror increases.
\begin{figure*}[h]
	\centering
	\includegraphics[width=\linewidth]{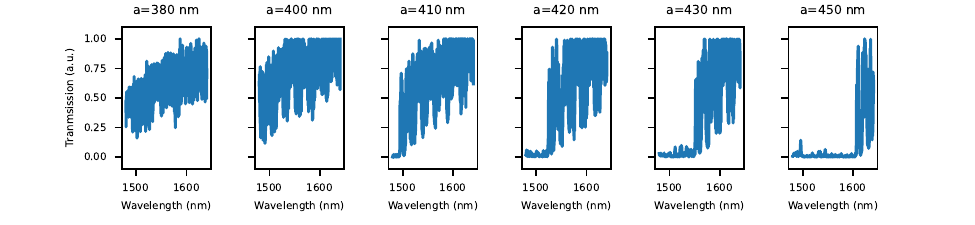}
	\caption{\textbf{Bragg mirror bandgap measurements.} {\bf a,} Transmission spectra of integrated TFLN Fabry-Perot cavities where the Bragg mirrors have a width of $w_w\equal 1000$~nm, a duty cycle of $d_a\equal0.8$, and periods ranging from $a\equal380$~nm to $a\equal450$~nm.}
        \figlab{round1Trans}
\end{figure*}
\begin{figure*}[h]
	\centering
	\includegraphics[width=\linewidth]{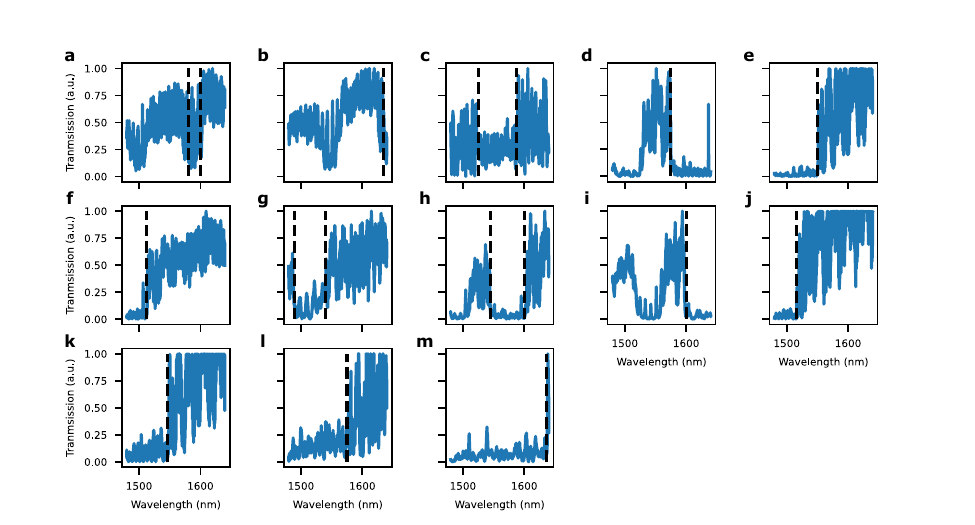}
	\caption{\textbf{Bragg mirror bandgap measurements for informing model parameters.} Transmission spectra of integrated TFLN Fabry-Perot cavities featuring prominent band edges. The design parameters ($d_a$ [\%], $a$ [nm], $w_w$ [nm]) for the devices are: \textbf{a,} (30, 430, 1000) \textbf{b,} (30,450, 1000) \textbf{c,} (50,430, 1000) \textbf{d,} (50, 450, 1000) \textbf{e,} (80, 430, 1000) \textbf{f,} (50, 400, 1500) \textbf{g,} (50, 410, 1500) \textbf{h,} (50, 430, 1500) \textbf{i,} (50, 450, 1500) \textbf{j,} (80, 410, 1500) \textbf{k,} (80, 420, 1500) \textbf{l,} (80, 430, 1500) \textbf{m,} (80, 450, 1500).}
        \figlab{round1Trans BGs}
\end{figure*}
\begin{figure*}[h]
	\centering
	\includegraphics[width=0.8\linewidth]{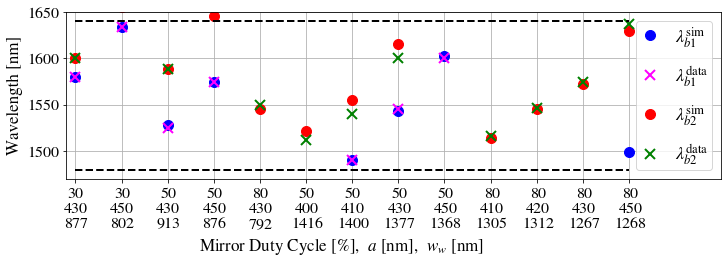}
	\caption{\textbf{Band-edge wavelengths comparison between model and experiment.} The experimental band edges are denoted $\lambda_{bn}^{\rm{data}}$ with $n$ representing the dielectric ($n\equal 1$) and air-like ($n\equal 2$) bands. Values extracted from measured transmission spectra of devices with varying grating duty cycles ($d_a$), periods ($a$), and waveguide widths $w_m$ are shown. The corresponding modeled band edges $\lambda_{bn}^{\rm{sim}}$ are calculated using MIT Photonic Bands (MPB). Our simulations used the geometrical parameters provided along the horizontal axis that were extracted through SEM imaging. The simulations assume an inner sidewall angle of $\theta_i=17^\text{o}$.}
        \label{fig:Round1 comparison}
\end{figure*}
%


Additional transmission spectra of similar cavities with mirrors having different design parameters are shown in~\figref{round1Trans BGs}, where we label the extracted location of the band edges with black dashed lines. We further validate geometric device features extracted from SEM imaging with numerical transmission spectra that have features similar to the experimental data. For instance, in Supplementary Figure~\ref{fig:Round1 comparison}, we provide plots of numerically calculated band edge positions for our device geometries extracted from SEM imaging. As further validated by the SEM images, a $17^\text{o}$ inner sidewall angle provides the best agreement with experiment.

\begin{figure*}[h]
	\centering
	\includegraphics[width=0.775\linewidth]{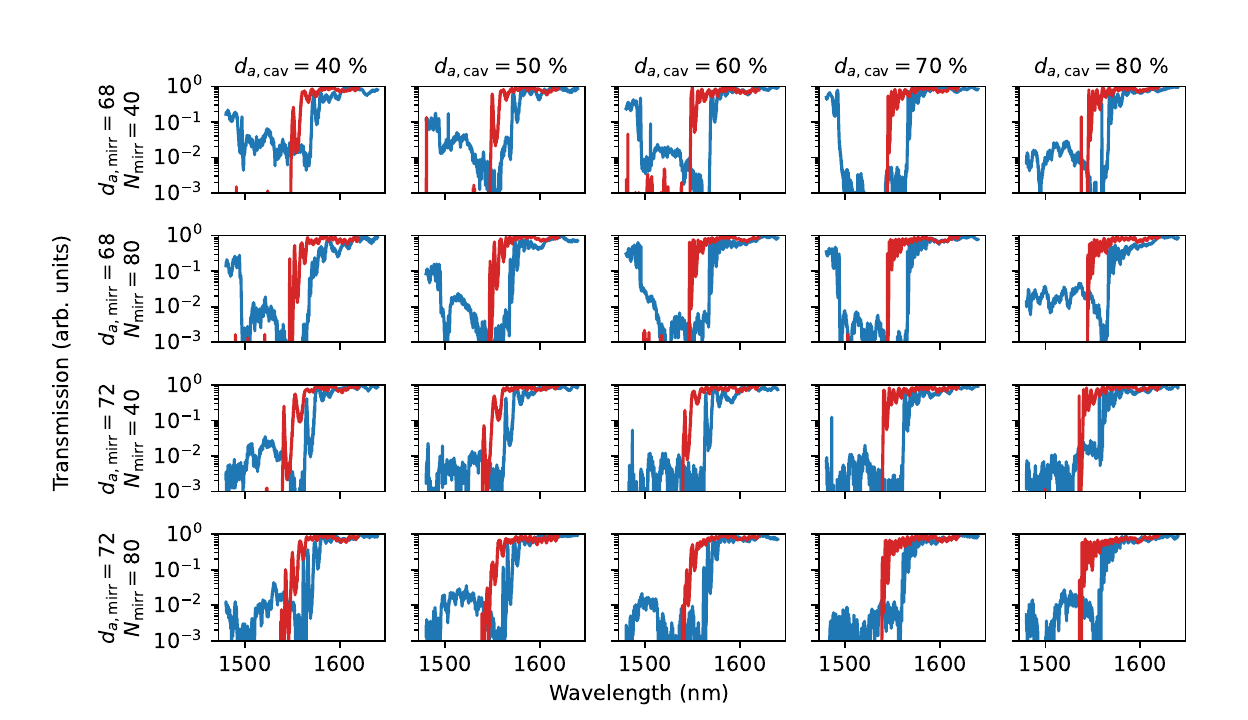}
	\caption{\textbf{Transmission spectra of photonic crystal cavities with a 422~nm period.} Transmission spectra for devices with Bragg mirror duty cycles, $d_{a,\text{mirr}}$, of 68\% and 72\% with $N_\text{mirr}=40$ and $N_\text{mirr}=80$ Bragg periods for devices with cavity regions formed by a Bragg grating with a duty cycle, $d_{a,\text{cav}}$, ranging  from 40\% to 80\%. Different plot colors correspond to devices with the same design parameters  fabricated on different wafers.}
    \label{fig:trans422}
\end{figure*}

\begin{figure*}[h]
	\centering
	\includegraphics[width=0.775\linewidth]{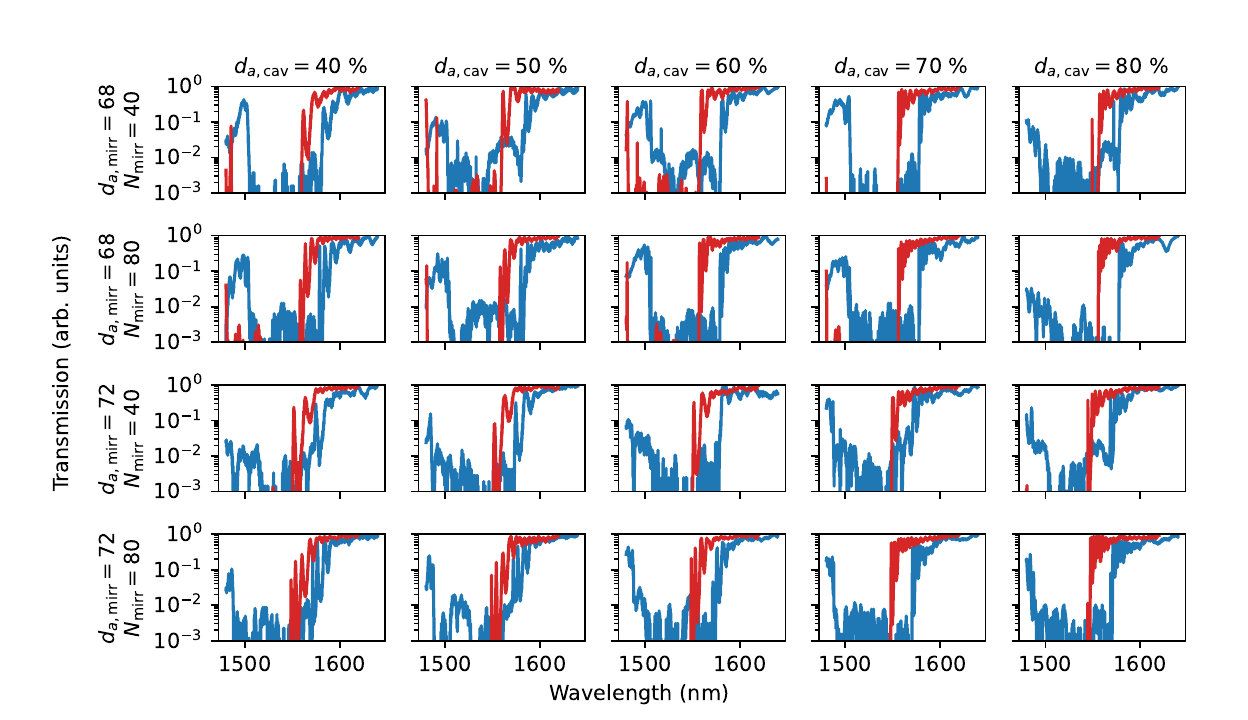}
	\caption{\textbf{Transmission spectra of photonic crystal cavities with a 426~nm period.} Transmission spectra for devices with Bragg mirror duty cycles, $d_{a,\text{mirr}}$, of 68\% and 72\% with $N_\text{mirr}=40$ and $N_\text{mirr}=80$ Bragg periods for devices with cavity regions formed by a Bragg grating with a duty cycle, $d_{a,\text{cav}}$, ranging  from 40\% to 80\%. Different plot colors correspond to devices with the same design parameters  fabricated on different wafers.}
        \label{fig:trans426}
\end{figure*}

\begin{figure*}[h]
	\centering
	\includegraphics[width=0.775\linewidth]{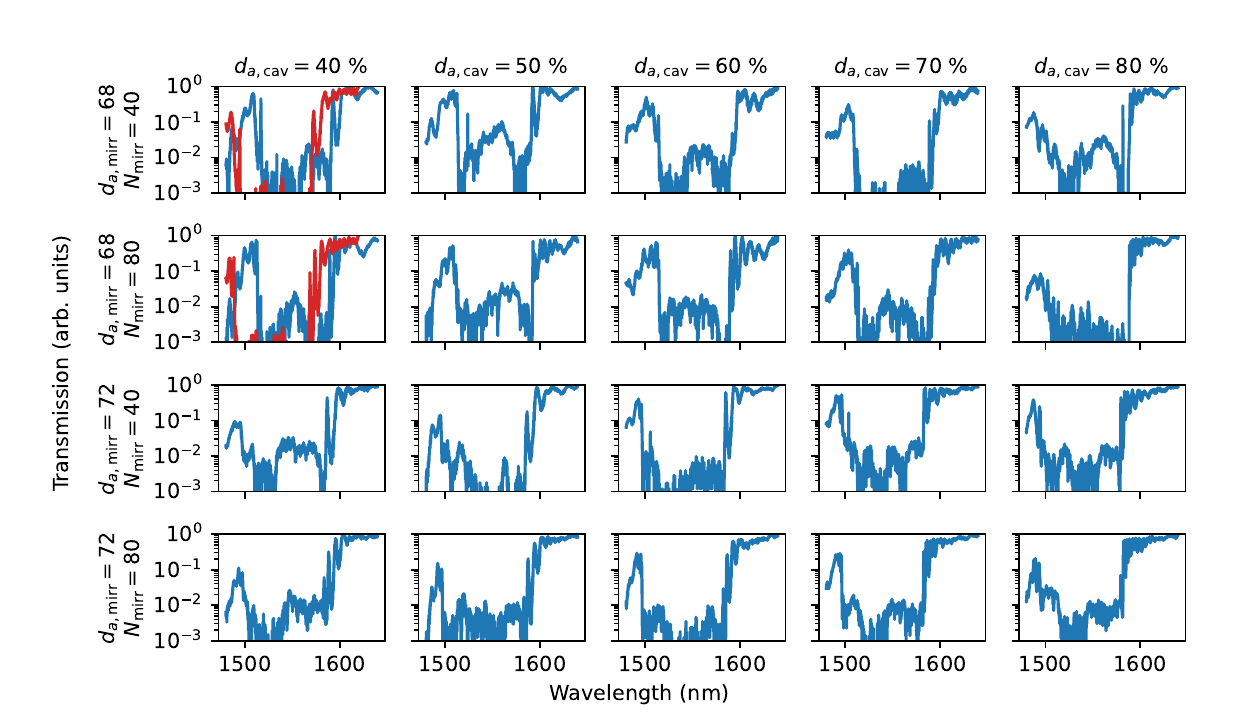}
	\caption{\textbf{Transmission spectra of photonic crystal cavities with a 430~nm period.} Transmission spectra for devices with Bragg mirror duty cycles, $d_{a,\text{mirr}}$, of 68\% and 72\% with $N_\text{mirr}=40$ and $N_\text{mirr}=80$ Bragg periods for devices with cavity regions formed by a Bragg grating with a duty cycle, $d_{a,\text{cav}}$, ranging  from 40\% to 80\%. Different plot colors correspond to devices with the same design parameters  fabricated on different wafers. Missing plots are attributed to devices that have been damaged from handling due to their proximity to the edge of the chip.}
     \label{fig:trans430}
\end{figure*}

Based on results from our first fabrication run, we updated the model parameters in subsequent fabrication iterations where Bragg mirrors all had duty cycles close to 70\% and periods near 425~nm. Supplementary Figures~\ref{fig:trans422},~\ref{fig:trans426}, and~\ref{fig:trans430} provide measured transmission spectra of such two-sided cavities with the geometry presented in~\figref{cavity parameters}a. All devices feature a bandgap in the targeted c-band region.
%

\subsection{Influence of Device Geometry on Transmission}
\begin{figure*}[h]
	\centering
\includegraphics[width=1\linewidth]{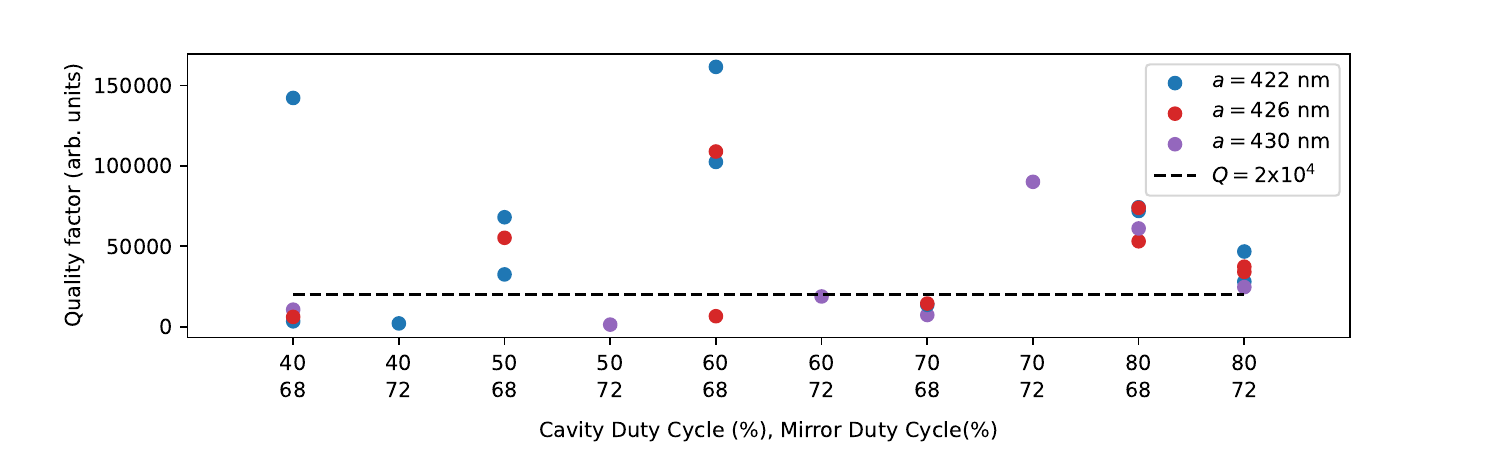}
	\caption{\textbf{Wafer-scale TFLN PhC cavity fabrication.} Loaded quality factors of resonances measured in two-sided TFLN PhC cavities with $N_{\rm{mirr}}\equal 40$. We also fabricated devices with $N_{\rm{mirr}}\equal 80$, however, due to the resulting high Bragg mirror reflectivity, we did not observe any resonances in their transmission spectra. Of the 51 measured $N_{\rm{mirr}}\equal 40$ devices, 28 featured resonances. Of these 28 devices, 18/28=64\% featured a loaded quality factor exceeding $2\times 10^4$, which was the threshold established in~\secref{PhC4QAM} for a PhC IQ modulator performing 4-QAM.}
    \label{fig:qFactors}
\end{figure*}

A set of design parameters yielded devices exhibiting visible resonances in their transmission spectra. In particular, we did not observe any resonances in cavities with  Bragg mirrors that have $N_\text{mirr}=80$ given that we expect their transmission to be quite low, thereby reducing coupling to the  cavity mode. Supplementary Figure~\ref{fig:qFactors} provides the loaded quality factors of double-sided cavities with $N_\text{mirr}=40$. Cavities with an 80~\% duty cycle in the cavity region yield consistent quality factors across devices of varying nominal periods on two wafers. 

\clearpage 
\section{Tuning Efficiency and Bit-Switching Energy}
To calculate the tuning efficiency, we represented our structure in COMSOL as illustrated in~\figref{simANDequivCirc}a and solved the Poisson equation with a fixed potential boundary condition at the electrode surfaces. Setting the voltages on the electrodes to 0 and $V_{\rm{DC}}\equal 1\,$V, we find $\partial_V\omega_n$ from~\eqref{resonance shift simple b} by inserting the corresponding DC field. The optical mode distribution of the cavity, $\vec{E}_n(\vec{r},\omega_n)$, was found from finite-difference-time-domain simulations using Ansys Lumerical and is shown in~\figref{cavity parameters}a. Using the values $\epsilon_o(\omega_{\rm{DC}})\equal 28$, $\epsilon_e(\omega_{\rm{DC}}) \equal 28$~\cite{Li:20},  $\epsilon_o(\omega_n)\equal 2.21^2$, $\epsilon_e(\omega_n) \equal 2.13^2$, $r_{13}\equal 9.6\,$pmV$^{-1}$, and $r_{33}\equal 30.9\,$pmV$^{-1}$~\cite{Zhu:21}, we find a tuning efficiency of $\partial_V\omega_n\equal\,$ $2\pi\times1.0$~GHzV$^{-1}$.\\

Modulator bit-switching energy consumption is estimated through analysis of an equivalent circuit model in an extension of Miller's approach \cite{Miller:12}. Based on the structure~\figref{simANDequivCirc}a, we use COMSOL to calculate the Maxwell capacitance matrix
\begin{equation}
    \begin{bmatrix}
        c_{11} & c_{12} \\
        c_{21} & c_{22}
    \end{bmatrix} = 
    \begin{bmatrix}
        20.0 & -9.4 \\
        -9.4 & 31.6
    \end{bmatrix} \textrm{fF},
\end{equation}
from which we calculate~\cite{COMSOLblog} the mutual capacitance values
\begin{equation} \label{eq:lumpedCapacitances}
    C_{11} = 10.6\, \textrm{fF}, C_{12} = C_{21} = 9.4 \, \textrm{fF}, \textrm{ and } C_{22} = 22.2 \, \textrm{fF},
\end{equation}
which corresponds to the lumped element model shown in~\figref{simANDequivCirc}{\bf b}.

We make a 0D time-dependent COMSOL simulation of the equivalent circuit and use it to calculate the energy dissipated in each resistor while transitioning between all possible modulator encoding settings with $V_1$ and $V_2$ each supplying 0-2 V (\tabref{transitionEnergies}). In addition to the transitions in~\tabref{transitionEnergies}, there are four zero-energy transitions where the modulator voltage settings do not change between adjacent transmitted symbols, and no energy is dissipated to change the state. Thus, assuming all symbols are equiprobable, the average energy dissipated in changing modulator states is 51.5 fJ/symbol, and since each symbol encodes 2 bits in 4-QAM, the average bit-switching energy is 25.8 fJ/bit. This can be further reduced with appropriate design modifications~\cite{Li:20}.
\begin{figure*}[th]
    \centering
    \includegraphics[width=\linewidth]{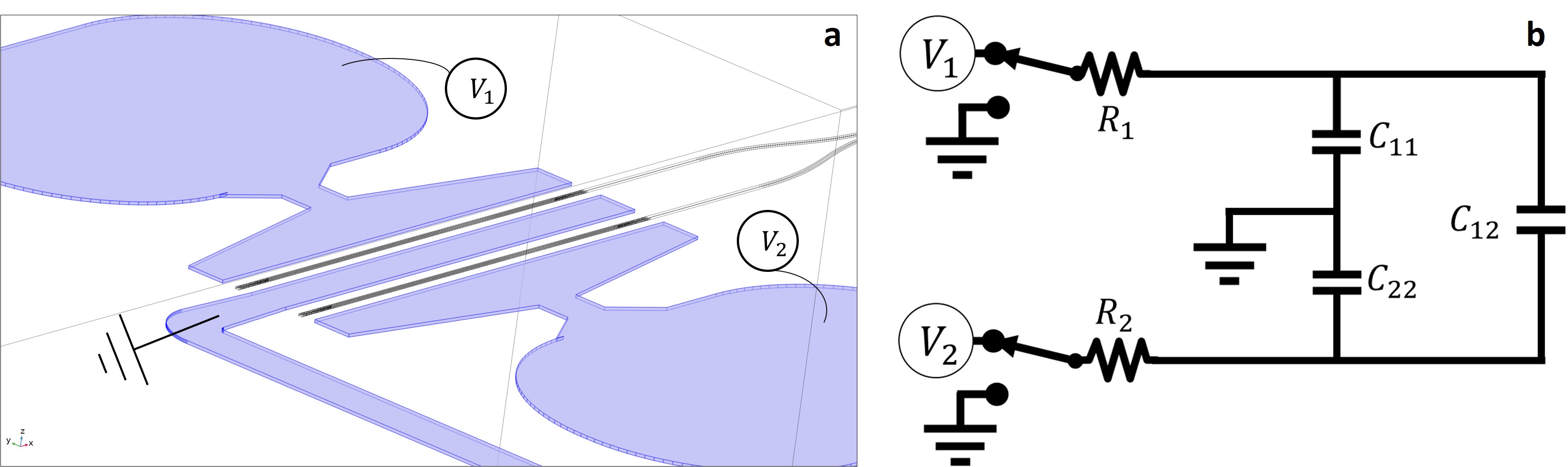}
    \caption{\textbf{Energy consumption simulations.} \textbf{a,} 3D electrostatic model of the TFLN PhC IQ modulator used to calculate the Maxwell capacitance matrix. \textbf{b,} Equivalent circuit model of the device where $C_{ij}$ refer to the mutual capacitance values of the system, whereas $R_n$ account for the in-series resistance between the voltage source $V_n$ and the device.}
    \figlab{simANDequivCirc}
\end{figure*}
\begin{table}
    \centering
    \begin{tabular}{|c|c|c|c|} 
    \hline
    \rowcolor{Gray}
      \bf Transition: $\bf (V_{1,i},V_{2,i}) \to (V_{1,f},V_{2,f})$ [V]  & \bf Energy dissipated in $\bf R_1$ [fJ] & \bf Energy dissipated in $\bf R_2$ [fJ] & \bf Total energy dissipated [fJ] \\ \hline 
        $(0,0) \to (2,0) $ & 36.7 & 3.4  & 40.1 \\ \hline
        \rowcolor{Gray}
        $(0,0) \to (0,2) $ & 3.4 & 59.9 & 63.3 \\ \hline
        
        $(0,0) \to (2,2) $ & 25.3 & 40.0  & 65.3 \\ \hline
        \rowcolor{Gray}
        $(0,2) \to (2,2) $ & 36.3 & 3.5  & 39.8 \\ \hline
        
        $(0,2) \to (2,0) $ & 54.7 & 86.4  & 141.1 \\ \hline
        \rowcolor{Gray}
        $(0,2) \to (0,0) $ & 3.5 & 59.6  & 63.1 \\ \hline
        
        $(2,0) \to (2,2) $ & 3.5 & 59.5 & 63.0 \\ \hline
        \rowcolor{Gray}
        $(2,0) \to (0,2) $ & 54.2 & 86.3  & 140.5 \\ \hline
        
        $(2,0) \to (0,0) $ & 36.4  & 3.5 & 39.9 \\ \hline
        \rowcolor{Gray}
        $(2,2) \to (0,2) $ & 36.3 & 3.5  & 39.8 \\ \hline
        
        $(2,2) \to (2,0) $ & 3.5 & 59.5  & 63.0 \\ \hline
        \rowcolor{Gray}
        $(2,2) \to (0,0) $ & 25.3  & 40.0 & 65.3 \\ \hline
    \end{tabular}
    \caption{{\textbf{Energy dissipated in transitions between modulator voltage states.}} We denote the initial (final) voltage on terminal $j\in \{1,2 \}$ as $V_{j,i}$ ($V_{j,f}$). The total energy dissipated is the sum of the energies dissipated in $R_1$ and $R_2$.}
    \tablab{transitionEnergies}
\end{table}
%

\section{DC Transmission Measurements and Model Fitting \seclab{DC Transmission Measurements and Model Fitting}}
Applying a potential to the electrodes on either side of each PhC cavity creates an electric field that changes the refractive index of the TFLN through the Pockels effect. Treating the interaction with the optical cavity field using first-order perturbation theory (see~\secref{Perturbation Theory}) results in the linear relationship:
\begin{align} \eqlab{EO model}
\omega_n = \omega_n^{(0)} + \frac{\partial \omega_n}{\partial V} V_n = \omega_n^{(0)} + \partial_V \omega_n V_n,
\end{align} 
where $V_n$ is the voltage across cavity $n$ and $\partial_V \omega_n$ is the tuning efficiency. Inserting this relation into~\eqref{r_n spectrum} allows us to define normalized parameters
\begin{align} \eqlab{cmt EO model}
    r_n(\omega) &= 1 - \frac{2\kappa_{c,n}}{2i(\omega_n - \omega) + \kappa_{n}} = 1 - 2\left( \frac{\kappa_{c,n}}{\kappa_{n}}\right) \frac{1}{ \displaystyle 2i\bigg[\frac{1}{\kappa_n} \Big(\omega_n^{(0)} + \partial_V\omega_n V_n \Big) - \frac{\omega}{\kappa_n}\bigg] + 1} = 1 - \frac{\displaystyle 2\left( \frac{x_n}{x_n + 1}\right)}{ 2i\Big(\tilde{\omega}_n^{(0)} - \tilde{\omega} + \partial_V \tilde{\omega}_n V_n\Big) + 1}, 
\end{align} 
where
\begin{align}\eqlab{app normalization defs}
    \tilde{\omega} = \frac{\omega}{\kappa_n}, ~~~\tilde{\omega}_n^{(0)} = \frac{\omega_n^{(0)}}{\kappa_n}, ~~~\partial_V \tilde{\omega}_n  = \frac{1}{\kappa_n} \frac{\partial\omega_n}{\partial V}, ~~\text{and} ~~ x_n = \frac{\kappa_{c,n}}{\kappa_{i,n}}. 
\end{align}
We perform two types of experiments to extract the model parameters that best describe our device. In the first experiment, we fix the wavelength of the incident CW laser near the resonance of the cavities and vary both voltages, $V_1$ and $V_2$. The setup used for these transmission measurements is illustrated in~\figref{transSetup}, and the corresponding 2D transmission map is shown in~\figref{TransFit}a, which is identical to Fig. 4b of the main text.
\begin{figure*}[h!]
    \centering
    \includegraphics[width=0.6\linewidth]{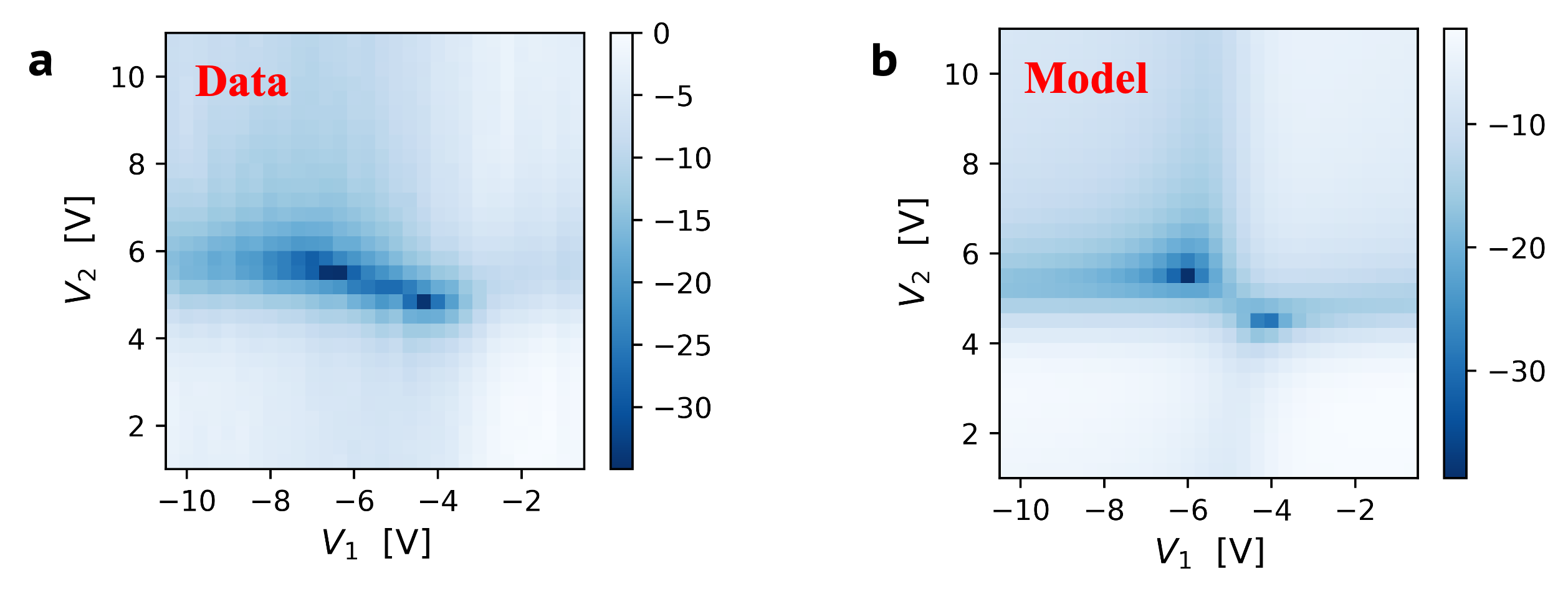}
    \caption{\textbf{Modulator transmission as a function of voltage.} {\bf a,} Experimental transmission maps of $|t_\text{IQ}(V_1, V_2)|^2$ (in dB). The data is normalized to a maximum value of 1. {\bf b,} Transmission maps calculated from~\eqsref{tIQ}{cmt EO model} using the parameters listed in~\tabref{model parameters}.}
    \figlab{TransFit}
\end{figure*}
Comparing the data to a transmission map found using~\eqsref{tIQ}{cmt EO model} allows us to find the fitting parameters: $\tilde{\omega}_n^{(0)}-\tilde{\omega}$, $\partial_V\tilde{\omega}_n$, $x_n$, and $\Delta\phi$. However, to find the quality factor, $Q_n\equal \omega_n/\kappa_n$, we must perform another set of experiments, where the laser wavelength is varied along with the voltage across only one of the cavities. 
\begin{figure*}[h!]
	\centering
	\includegraphics[width=0.9\linewidth]{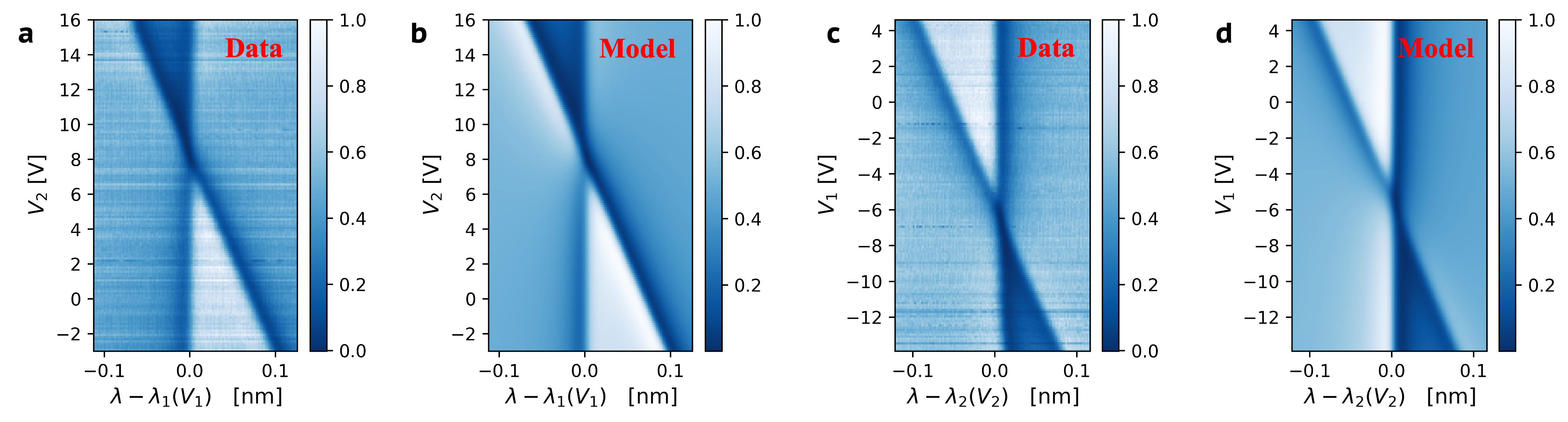}
	\caption{\textbf{Modulator transmission as a function of wavelength and voltage.} {\bf a,} Experimental transmission maps of $|t_\text{IQ}(\lambda, V_2)|^2$, normalized to a maximum value of 1. {\bf b,} Model prediction of $|t_\text{IQ}(\lambda, V_2)|^2$ calculated from~\eqsref{tIQ}{cmt EO model} using the parameters listed in~\tabref{model parameters}. {\bf c,} Experimental transmission maps of $|t_\text{IQ}(\lambda, V_1)|^2$, normalized to a maximum value of 1. {\bf d,} Model prediction of $|t_\text{IQ}(\lambda, V_1)|^2$ calculated from~\eqsref{tIQ}{cmt EO model} using the parameters listed in~\tabref{model parameters}.}
	\label{fig:SpecFits}
\end{figure*}
Two examples of such 2D transmission maps are shown in~\figref{SpecFits}a,c, where $V_1$ and $V_2$ are held fixed, respectively. Note that these plots are identical to Fig. 3b,c in the main text, except a linear scale is used for the transmission here. In both plots, one transmission dip appears at a fixed wavelength while the other shifts linearly with voltage, which matches the model in~\eqref{EO model}. The slope of the linear line through the transmission minima yields the value of $\partial_V\omega_n$, and $\kappa_n$ is found by comparing to the normalized value $\partial_V\tilde{\omega}_n$. The decay rate, $\kappa_n$, may also be estimated by comparing the transmission calculated from~\eqref{cmt EO model} to single wavelength scans at voltage levels where the resonances are well-separated. The model parameters that best fit the measured data are found by minimizing
\begin{align}\eqlab{def fit error}
   \sum_\omega\sum_{V_1}\sum_{V_2} \Big| \big|t_{\rm{IQ}}^{\rm{data}}(\omega,V_1,V_2)\big|^2 - \big|t_{\rm{IQ}}^{\rm{model}}(\omega,V_1,V_2)\big|^2 \Big|
\end{align}
using all the data in~\figsref{TransFit}{SpecFits}, and the extracted parameters are listed in~\tabref{model parameters}.
\begin{table}[!h]
\renewcommand{\arraystretch}{1.4}
\vspace{3mm}
   \small
    \begin{tabular}{ |p{1.5cm}|p{2.5cm}|p{1.5cm}|p{2.5cm}|  }
    \hline
    \rowcolor{Gray}
     \multicolumn{2}{|c|}{\bf Cavity 1} & \multicolumn{2}{c|}{\bf Cavity 2} \\
     \hline
     \makecell{$\lambda_1^{(0)}$}  & \makecell{1548.60$\,$nm}  & \makecell{$\lambda_2^{(0)}$}    & \makecell{1548.71$\,$nm}  \\
     \hline
     \rowcolor{Gray}
     \makecell{$Q_1$} & \makecell{$6.6 \times 10^4$} & \makecell{$Q_2$} & \makecell{$7.4 \times 10^4$} \\
     \hline
     \makecell{$\kappa_1$} & \makecell{$2\pi \times  2.9$~GHz} & \makecell{$\kappa_2$} & \makecell{$2\pi \times  2.6$~GHz} \\
     \hline
     \rowcolor{Gray}
     \makecell{$\kappa_{c,1}/\kappa_{i,1}$} & \makecell{0.46} & \makecell{$\kappa_{c,2}/\kappa_{i,2}$} & \makecell{1.42} \\
     \hline
     \makecell{$\partial_V\omega_1$} & \makecell{$2\pi\times 1.15$~GHz/V} & \makecell{$\partial_V\omega_2$} & \makecell{$2\pi\times 1.09$~GHz/V} \\
     \hline
     \rowcolor{Gray}
     \multicolumn{4}{|c|}{\bf Michelson Interferometer}  \\
     \hline
     \makecell{$\Delta\phi$} & \makecell{$0.63\pi$} & \makecell{$\zeta$} & \makecell{$\sqrt{0.12}$} \\
     \hline
    \end{tabular}
    \caption{Model parameters extracted by fitting the measured data in~\figref{TransFit}a and~\figref{SpecFits}a,c using~\eqref{tIQ}. }
    \tablab{model parameters}
\end{table}
%

Transmission maps calculated from~\eqsref{tIQ}{cmt EO model} using the parameters in~\tabref{model parameters} are plotted in~\figref{TransFit}b and~\figref{SpecFits}b,d. A good agreement with the measured data is observed, providing confidence in the applicability of the coupled mode theory model in~\secref{Coupled Mode Theory Modeling}.

\section{Apparatus for Transmission Measurements }

\subsection{Intensity Transmission Apparatus\seclab{transApparatus}}

We provide a sketch of our apparatus for transmission measurements in~\figref{transSetup}. To measure the bandwidth of the IQ modulator's cavities, we replace the DC Sources with a vector network analyzer (Keysight N5224A) and our DC-coupled photodiodes with a fast AC-coupled detector (Thorlabs RXM10BF).
\begin{figure*}[h!]
	\centering
	\includegraphics[width=0.9\linewidth]{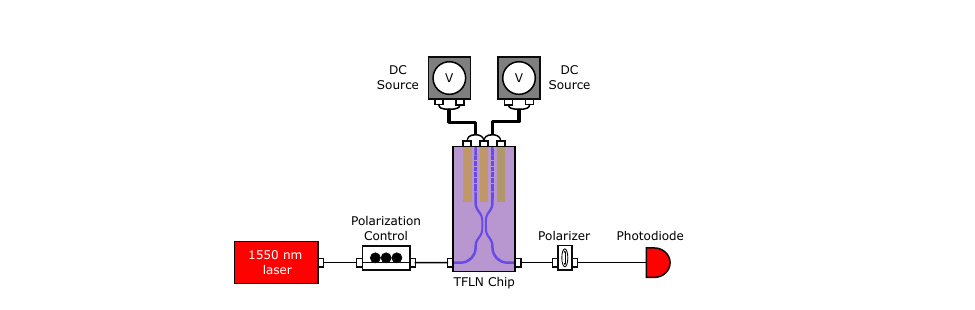}
	\caption{\textbf{Transmission measurement apparatus.} A tunable continuous-wave semiconductor cavity laser (Santec TSL-710) sends coherent light to the TFLN device under test. We rely on polarization control with polarization paddles and edge-coupling with 2.5~\textmu m lensed fibers (OZ Optics) to couple light into the TE mode of our TFLN waveguides. Electronic control of our devices is based on a source measure unit (Keithley 2400) and RF probes (Formfactor ACP series). The modulated optical field couples out of the chip via another edge coupler before going through a rotatable polarizer mounted in a fiber-to-fiber U-bench to filter out any polarization cross-talk resulting from on-chip propagation. The filtered signals are sent to a photodiode detector (Agilent 8163A) connected to a data acquisition module (NI USB-6259).}
        \figlab{transSetup}
\end{figure*}
%

\subsection{Coherent Transmission Apparatus \seclab{Apparatus for coherent measurements}}
We provide a sketch of our apparatus for coherent measurements in~\figref{coherentSetup}. The figure illustrates the setup's configuration while running the quadrature amplitude modulation (QAM) experiments from the main text. We also performed slower coherent measurements with this apparatus by replacing the pulse pattern generator with a 60~MHz arbitrary waveform generator (BK Precision 4055B) and a 100~MHz oscilloscope (BK Precision 2190E).

\begin{figure*}[h]
	\centering
	\includegraphics[width=0.9\linewidth]{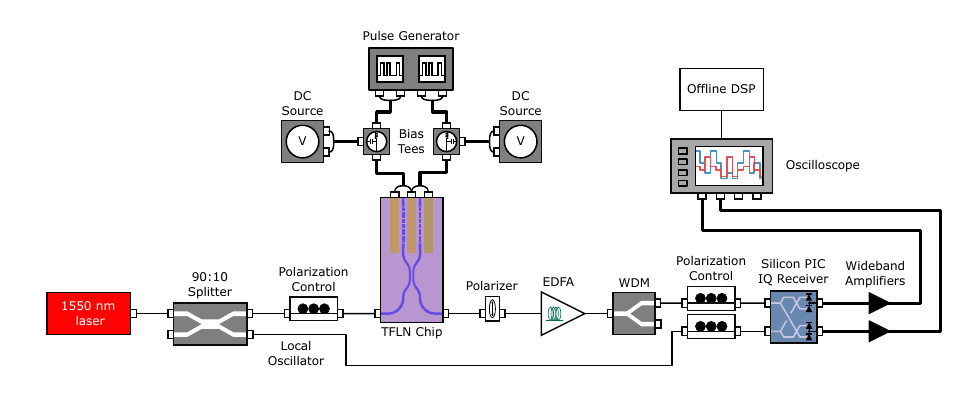}
	\caption{\textbf{Coherent apparatus.} A tunable continuous-wave semiconductor cavity laser (Santec TSL-710) sends coherent light to a 90:10 fiber splitter. The 10\% output connects to a polarization control unit before it is connected to the TFLN PhC IQ modulator via edge couplers. The 90\% output serves as a local oscillator input to the PIC IQ receiver. A 12.5 GHz pulse pattern generator (Anritsu MP1763B) sends pseudo-random bit sequences for our QAM experiments. The bandwidth of the resulting sequences is set to a quarter of the generator's bandwidth in order to use two of the generator's outputs to drive our cavity pair. We bias the bit sequences with bias tees (Minicircuit ZFBT-6GW+) and DC signals provided by a DC power supply (BK Precision 9129B) to align the resonances of the IQ modulator's cavities. RF probes (Formfactor ACP series) route the resulting biased electrical signals to the TFLN chip. The resulting EO-modulated optical field couples out of the chip via another edge coupler before going through a rotatable polarizer mounted in a fiber-to-fiber U-bench to filter out any polarization cross-talk resulting from on-chip propagation. An Erbium-doped fiber amplifier (Oprel OFA17D-12141M) amplifies the circuit signal, followed by a wavelength division multiplexing filter (Fiberdyne Labs) to remove excess amplified spontaneous emission noise. The modulated signal and the local oscillator are sent through polarization control units before being coupled into a custom-made silicon photonic integrated circuit (PIC) IQ receiver by means of edge coupling using a v-groove fiber array (OZ Optics). Photocurrents attributed to the $I$ and $Q$ quadratures of the measured optical field are sent through 2~GHz amplifiers (FEMTO HSA-X-2-40) before finally reaching a 10~GHz oscilloscope (Infiniium DSO81004A) for measurements.}
    \figlab{coherentSetup}
\end{figure*}

 To measure the $I$ and $Q$ quadratures of our modulated optical fields, we relied on a custom-built silicon photonic integrated circuit (PIC) IQ receiver manufactured by AIM photonics in a multi-project wafer run. We provide a micrograph of the corresponding device in~\figref{iqReceiver}. 
 \begin{figure*}[h!]
	\centering
	\includegraphics[width=0.9\linewidth]{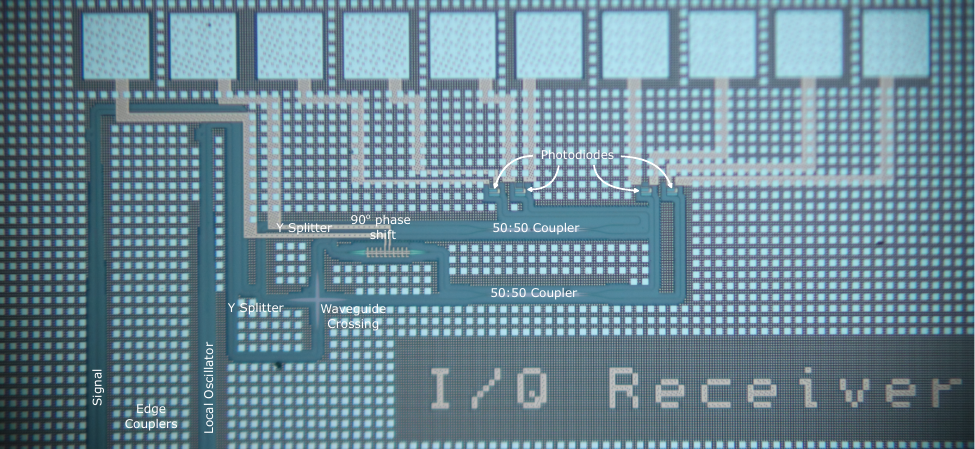}
	\caption{\textbf{Silicon photonic integrated circuit IQ receiver.} Edge couplers transmit a modulated optical signal and a local oscillator from cleaved SMF-28 fibers into the PIC. The two fields are fed into a $90^\text{o}$ hybrid consisting of Y-splitters followed by a thermo-optic phase shifter configured to impart a $90^\text{o}$ phase shift, and finally, a pair of 50:50 couplers. The PIC routes the hybrid's output to on-chip photodiodes sending the resulting photocurrents out of the chip via pads wire-bonded to an RF printed circuit board (PCB).}
        \figlab{iqReceiver}
\end{figure*}
 We send our modulated optical signal and a local oscillator to the PIC via edge couplers and then route them to a $90^\text{o}$ hybrid, which involves splitting them and performing two homodyne measurements on the signal: One with the local oscillator and the other with a $90^\text{o}$ phase shifted version of it. We impart the $90^\text{o}$ phase shift by means of a thermo-optic phase shifter. The resulting interfered optical fields are sent to on-chip photodetectors with a bandwidth of 30~GHz. The generated electrical signals are sent to off-chip amplifiers before reaching an oscilloscope. We provide the performance metrics of the various components of this IQ receiver in Supplementary Table~\ref{table:iqReceiver}.
\begin{table}[h]
\centering
\begin{tabular}{ |p{4cm}|p{4cm}|p{4cm}| }
\hline \rowcolor{Gray}
 \thead{\bf Component}  & \thead{\bf Metric} & \thead{\bf Performance}\\
 \hline
 \makecell{Edge coupler} & \makecell{Insertion Loss (dB)} & \makecell{$\leq 3$~dB} \\ \hline
 \rowcolor{Gray}
 \makecell{Y-splitter} & \makecell{Insertion Loss (dB)} & \makecell{$\leq 0.5$~dB} \\ \hline
 \makecell{Waveguide crossing} & \makecell{Insertion Loss (dB)} & \makecell{$\leq 0.25$~dB} \\ \hline
 \rowcolor{Gray}
 \makecell{Phase shifter} & \makecell{Insertion Loss (dB)} & \makecell{$\leq 0.25$~dB} \\ \hline
 \makecell{50:50 coupler} & \makecell{Insertion Loss (dB)} & \makecell{$\leq 0.5$~dB} \\ \hline
 \rowcolor{Gray}
 \makecell{Photodetectors} & \makecell{Responsivity} & \makecell{1 A/W} \\ \hline
 \makecell{Amplifiers} & \makecell{Transimpedance gain} & \makecell{5000 V/A} \\ 
 \hline
\end{tabular}
\caption{\textbf{IQ Receiver Performance.} Performance metrics of the individual on- and off-chip components forming the IQ receiver used for coherent measurements.}
\label{table:iqReceiver}
\end{table}
From these performance metrics, we expect the receiver to have a detection efficiency of 1769~V/W. In practice, we measure an efficiency going up to 77\% of the value expected from the specifications of each component.

To calibrate the $90^\text{o}$ phase shift, we replace our TFLN chip with a commercial lithium niobate EO phase modulator (Thorlabs LN65S-FC) and drive it with a sawtooth waveform shown in~\figref{phaseCal}a with an amplitude corresponding to the modulator's $V_\pi$ voltage of around 8~V. This modulation effectively adds a linear phase ramp to the output field sent to the IQ receiver, which manifests as $90^\text{o}$ offset sinusoids in the field's two quadratures, i.e.
\begin{equation}
    \text{Modulated field} \propto \left[ e^{i f t} \right]= \left[ \cos(ft) + i \sin(ft)\right] = \left[ \cos(ft) + i \cos(ft+\pi/2)\right],
\end{equation}
where $f$ is the frequency of the driving waveform, whereas the first and second terms of the last expression correspond to the $I$ and $Q$ quadratures of the modulated field, respectively. We sweep the voltage applied to the receiver's thermo-optic phase shifter and extract the phase offset between the sinusoids measured in the two channels of the receiver, such as the ones shown in~\figref{phaseCal}b, with Fourier analysis. The results of this calibration procedure are shown in~\figref{phaseCal}c, which indicates that 3.3~V induces the required $90^\text{o}$ delay required to use the PIC as an IQ receiver.
\begin{figure*}[h]
	\centering
	\includegraphics[width=\linewidth]{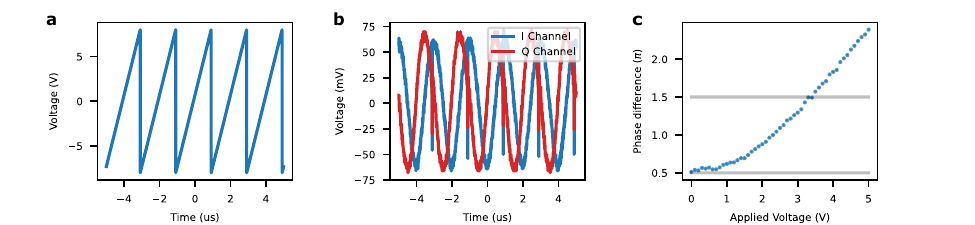}
	\caption{\textbf{IQ receiver calibration.} {\bf a,} Sawtooth waveform driving the commercial lithium niobate EO phase shifter during the calibration of the IQ receiver. {\bf b,} Measured waveforms from the IQ receivers balanced photodiodes for an input signal modulated with the voltage in {\bf a}. {\bf c,} Relative phase between the sinusoids of the type shown in {\bf b} as a function of the voltage applied onto the IQ receiver's thermo-optic phase shifter.}
        \figlab{phaseCal}
\end{figure*}

As implied by~\figref{coherentSetup}, the off-chip amplifiers placed between the IQ receiver and the oscilloscope have the smallest bandwidth (2~GHz) and thereby limit the maximum detection bandwidth of our setup to this value. To verify this, we replace the TFLN chip in~\figref{coherentSetup} with the aforementioned commercial lithium niobate phase modulator (bandwidth of 10~GHz), drive it with pseudo-random bit sequences, and monitor the calibrated outputs of the IQ receiver on an oscilloscope.~\figref{response} illustrates the falling edge of a bit flip observed in the waveform. 
\begin{figure*}[h!]
	\centering
	\includegraphics[width=0.5\linewidth]{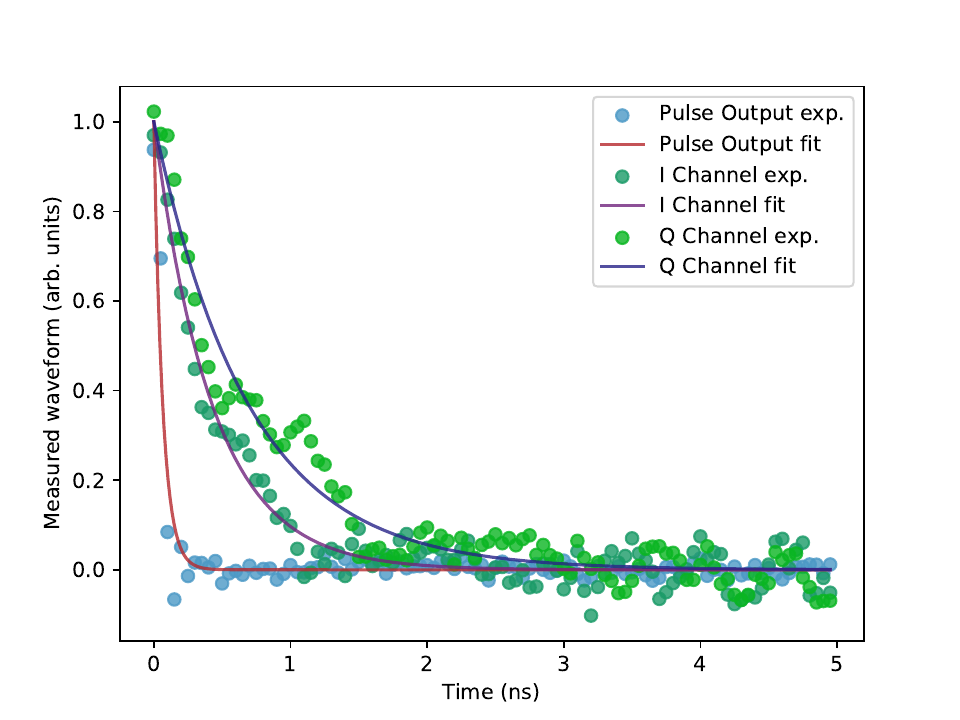}
	\caption{\textbf{Measurement apparatus rise time for IQ measurements.} Recorded samples on a 10~GHz oscilloscope featuring the voltage measured during a bit flip in a pseudo-random bit sequence. ``Pulse  Output'' refers to measured samples when the pulse pattern generator directly connects to the oscilloscope. ``I Channel'' and ``Q Channel'' data refer to recorded samples attributed to the amplified outputs of the IQ receiver when fed with light modulated by a commercial lithium niobate phase shifter driven by the pulse pattern generator.}
    \figlab{response}
\end{figure*}
Fitting this transition to a decaying exponential of the form $A \exp({-t/\tau})$, where $A$ is an amplitude scaling factor and $\tau$ is a time constant, yields time constants of $0.43 \pm 0.02$~ns and $0.69 \pm 0.03$~ns for the $I$ and the $Q$ channels, respectively, which are in-line with the expected 2~GHz bandwidth limit. As a reference,~\figref{response} also shows the transition between bits when the pulse pattern generator directly feeds into the oscilloscope. In this configuration, the fitted rise time is $0.065 \pm 0.004$~ns, which is close to the value expected from the 10~GHz limit of the oscilloscope.

~\figref{PICpictures} provides photographs of the TFLN IQ modulator and the silicon PIC IQ receiver in their respective parts of the measurement apparatus.

\begin{figure*}[h]
	\centering
	\includegraphics[width=\linewidth]{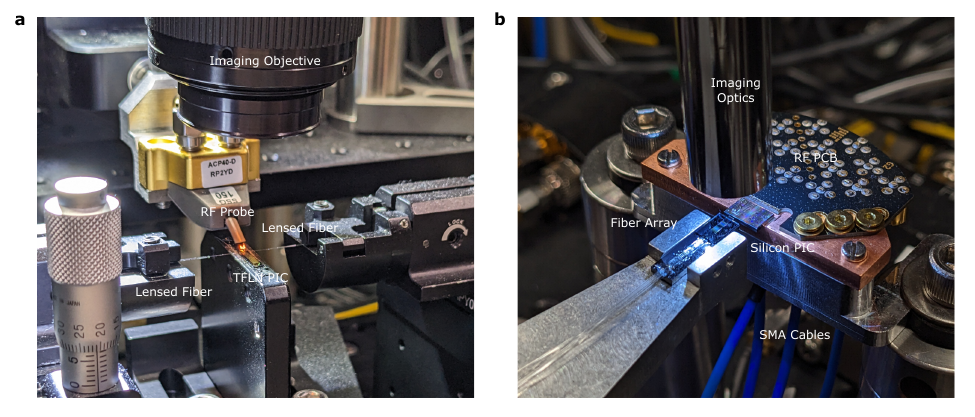}
	\caption{\textbf{Photographs of the photonic integrated circuits used in coherent transmission experiments.} {\bf a,} TFLN PhC IQ modulator and associated components. {\bf b,} Silicon PIC IQ receiver and associated components. {\bf Figure legend:} TFLN: thin-film lithium niobate, PIC: photonic integrated circuit, RF: radio frequency, PCB: printed circuit board.}
    \figlab{PICpictures}
\end{figure*}

\subsection{Coherent Transmission Measurements \seclab{Coherent transmission measurements}}

\begin{figure*}[h]
	\centering
	\includegraphics[width=\linewidth]{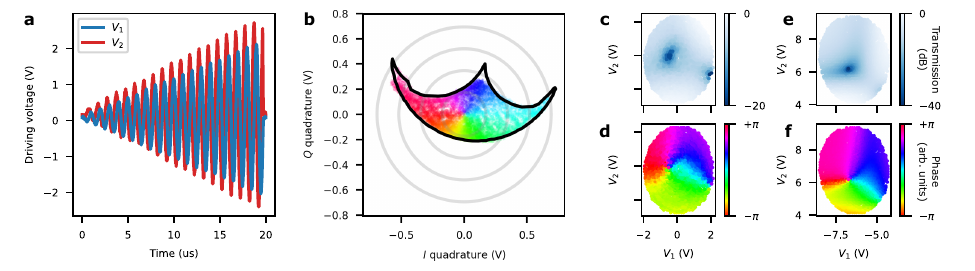}
	\caption{\textbf{Coherent transmission map measurements.} {\bf a,} $(V_1, V_2)$ voltages driving the IQ modulator for transmission map measurements. {\bf b,} Measured IQ quadratures while driving the IQ modulator with the waveforms in a. The drawn outline corresponds to that of the modeled IQ data. {\bf c,} $|t_\text{IQ}(V_1, V_2)|^2$ and {\bf d,} $\text{arg}(t_\text{IQ}(V_1, V_2))$ maps reconstructed from the data shown in {\bf b}. {\bf e,} Theoretical $|t_\text{IQ}(V_1, V_2)|^2$ and {\bf f,} $\text{arg}(t_\text{IQ}(V_1, V_2))$ maps obtained from model parameters.}
         \label{fig:coherentTrans}
\end{figure*}

In our coherent measurement experiments, the optical field going through the IQ modulator goes through the TFLN chip, whereas the local oscillator goes through a fiber patch cable. These different paths introduce a temporal phase drift affecting the measurement of the modulated field's two quadratures due to different environmental effects affecting the optical path length of each field. Though active feedback or additional monitoring could overcome these effects, we opt for a different method to construct $t_\text{IQ}(V_1,V_2)$ transmission maps, similar to the ones obtained from the setup shown in Supplementary Figure~\ref{fig:transSetup}, based on data collected from the IQ receiver. 

Our method involves driving the modulator with a pair of waveforms produced by an arbitrary waveform generator that covers a given region of the $(V_1, V_2)$ parameter space. As shown in Supplementary Figure~\ref{fig:coherentTrans}a, we rely on two sinusoids offset by $\pi/2$ and with linearly increasing amplitudes, which effectively trace out spirals in $(V_1,V_2)$ space. If the period of this waveform is faster than the setup's phase drift, then temporally aligning the measured IQ quadratures to these driving voltages based on the identical period of the two sets of waveforms allows us to construct  $t_\text{IQ}(V_1,V_2)$ transmission maps over a range limited by the maximum peak-to-peak voltage of the arbitrary waveform generator. Though this range can seem restricted, the high $Q$ factor and $\partial_V \omega$ value of the modulator's cavities ensure that relevant transmissive features fall within an accessible voltage range. Supplementary Figure~\ref{fig:coherentTrans}b provides the raw IQ quadrature data of the corresponding modulated field while biasing the driving waveforms of Supplementary Figure~\ref{fig:coherentTrans}a to one of the local minima of $|t_\text{IQ}(V_1,V_2)|^2$. Herein, the black curve corresponds to the outline of the points corresponding to theoretical data based on model parameters, which is in good agreement with the outline formed by the experimental data. Supplementary Figures~\ref{fig:coherentTrans}c,d plot the transmission maps $|t_\text{IQ}(V_1,V_2)|^2$ and $\arg\{t_\text{IQ}(V_1,V_2)\}$ extracted from the raw IQ data. We plot the corresponding data expected from our model parameters in Supplementary Figure~\ref{fig:coherentTrans}e,f. The experimental data exhibits quintessential transmission features expected from the model, which include the two local minima attributed to perfect destructive interference between the outputs of the two cavities, phase vortices located at these minima, and branches of low transmission going through them. We attribute the lower extinction in our phase measurements to the lower dynamic range of our coherent detection system and to the fact that we removed the polarizer filtering residual transmission due to polarization cross-talk in the TFLN chip in order to get stronger signals from the IQ receiver. Minor distortions in the experimental data are likely caused by a temporal offset between the measured $(V_1,V_2)$ and $t_\text{IQ}(V_1,V_2)$ due to slightly different acquisition settings.




\section{PhC Requirements for 4-QAM \seclab{PhC4QAM}}

When modulation voltages are constrained within a certain range, the linewidths of the modulator's PhC cavities ultimately dictate the dynamic range of the device along with other factors, such as insertion loss. Measurement noise, in turn, determines how many symbols can fit within this range. To establish a measure of the fabrication yield of our cavities, we determine the minimum quality factor required to run the 4-QAM experiments reported in this work. We assume that the laser used in our experiment outputs 12~mW of power, where 85\% goes to the local oscillator and 3\% to the TFLN modulator as determined by the insertion loss and splitting ratio of the employed fiber directional coupler shown in Supplementary Figure~\ref{fig:coherentSetup}. We also assume that the insertion loss at the facets of the TFLN chip is -8.2~dB (15\% transmission) as established from experimental measurements. This figure is limited by a faulty facet etch that occurred during the chip's fabrication. However, other fabrication runs suggest that this figure can reach -3~dB (50\% transmission). We then assume additional loss due to the 12:88 splitting ratio of the IQ modulator's directional coupler, as reported in Supplementary Figure~\ref{fig:directional_coupler}. The rest of the conversion metrics follow those reported in Supplementary Table~\ref{table:iqReceiver}. The noise in the experiment corresponds to the dark current of our integrated photodiodes, which is specified to 25~nA, and of the thermal noise of our wideband amplifiers, specified to 12.4~pA/$\sqrt{\text{Hz}}$.

\begin{figure*}[h]
    \centering
    \includegraphics[width=\linewidth] {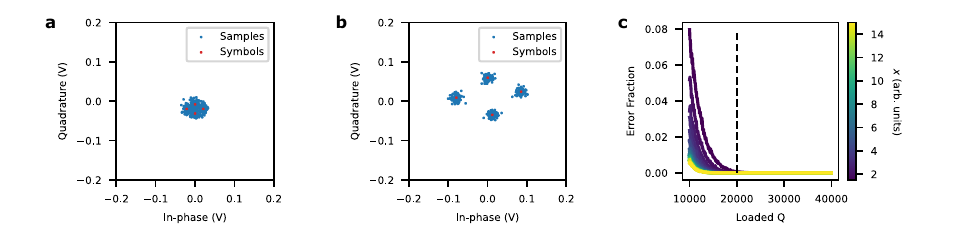}
    \caption{\textbf{Symbol overlap} \textbf{a,} Simulated constellation diagram with similar loss and noise figures as the ones in our experiments for an IQ modulator with cavity parameters of $\Delta \phi =0$, $x=2$, and $Q=10^4$, and \textbf{b,} $\Delta \phi =0$, $x=2$, and $Q=5 \times 10^4$. \textbf{c,} Fraction of misclassified samples for IQ modulators with various design parameters with the same noise figures as the ones in a and b. These results assume $\Delta \phi =0$ and are based off of a total of 400 samples.}
    \label{fig:qOpt}
\end{figure*}

Based on these settings, we numerically recreate constellation diagrams for a given cavity design based on~\eqref{tIQ}. We then run a constrained ADAM optimizer on the voltage configuration of each symbol that minimizes the overlap integrals of the underlying distributions responsible for the noise in the samples attributed to each symbol, which we assume to be normal for illustrative purposes. For example, Supplementary Figure~\ref{fig:qOpt}a provides the optimized constellation diagram in a modulator design that assumes two identical cavities with $\Delta \phi =0$, $Q=10^4$, and $x=2$. Supplementary Figure~\ref{fig:qOpt}b shows the corresponding data when we raise the quality factors of the cavities to $Q=5 \times 10^4$, thereby showing the influence of the narrow linewidths in the transmission of the device. In Supplementary Figure~\ref{fig:qOpt}c, we plot the fraction of misclassified samples, out of a total of 400, based on their proximity to the constellation diagram's symbols for various cavity designs, where we assume $\Delta\phi=0$ for convenience. We notice that this fraction zeroes out for $Q>20,000$, thereby suggesting this value as a cutoff to determine cavity fabrication yield. Based on the quality factors reported in Supplementary Figure~\ref{fig:qFactors}, we calculate this yield to be 64.3\%.

\bibliographystyle{naturemag}
\bigbreak
\def\bibsection{}  
\noindent\textbf{Supplementary References}
\bigbreak
\bibliography{phcBib}